\documentclass[11p]{article}
\usepackage[dvipdfmx]{graphicx}
\newcommand{\pf}{\mathrm{pf}}
\newtheorem{proposition}{Proposition}[section]
\newtheorem{theorem}{Theorem}[section]
\usepackage{color, amsmath}
\title{Discrete and Ultradiscrete Periodic Phase Soliton Equations}
\author{
Hidetomo Nagai$^1$\thanks{hdnagai@tokai-u.jp}, Yasuhiro Ohta$^2$\thanks{ohta@math.sci.kobe-u.ac.jp}, and Ryogo Hirota$^3$\thanks{Deceased 17 January 2015} 
}
\date{$^1$Department of Mathematics, Tokai University, 4-1-1 Kitakaname, Hiratsuka, Kanagawa 259-1292, Japan \\
$^2$Department of Mathematics, Kobe University, 1-1 Rokkodai, Nada, Kobe, Hyogo 657-8501, Japan \\
$^3$Professor Emeritus, Waseda University, 3-4-1 Okubo, Shinjuku, Tokyo 169-8555, Japan
}
\begin{document}
\maketitle 

\begin{abstract}
We propose new type of discrete and ultradiscrete soliton equations, which  admit extended soliton solution called periodic phase soliton solution.  The discrete equation is derived from the discrete DKP equation and the ultradiscrete one is obtained by applying the ultradiscrete limit.  The soliton solutions have internal freedom and change their shape periodically during propagation.  In particular, the ultradiscrete solution reduces into the solution to the ultradiscrete hungry Lotka-Volterra equation in a special case.    
\end{abstract}

\section{Introduction}
The periodic phase soliton (PPS) is a recently discovered wave for the ultradiscrete hungry Lotka-Volterra (uhLV) equation\cite{Nakamura}.  Usual soliton is a traveling wave which maintains its shape during propagation, while the PPS is the one which keeps changing its shape periodically.  Since the PPS solutions have many internal freedoms, they can exhibit various types of interactions.  As far as the authors know, this type of periodically oscillating solution has been given for the uhLV equation only.  The PPS in Ref. \cite{Nakamura} is the ultradiscrete solution and even its discrete analogue has not been reported yet.  \par
  In this paper the PPS solutions are constructed for the following discrete soliton equation,  
\begin{equation}  \label{nonlineardPPS}
\begin{aligned}
  \frac{u^{l+1}_{m}}{u^{l}_{m}}=& \prod _{k=1}^M \frac{1+\delta u^{l}_{m-k}}{1+\delta u^{l+1}_{m+k}}\frac{1+\delta w^{l+1}_{m+k}}{1+\delta w^{l}_{m-k}}, \\
  \frac{w^{l+1}_{m}}{w^{l}_{m}}=& \prod _{k=1}^M \frac{1+\delta u^{l}_{m-k+1}}{1+\delta u^{l+1}_{m+k-1}}\frac{1+\delta w^{l+1}_{m+k-1}}{1+\delta w^{l}_{m-k+1}},   
\end{aligned}
\end{equation}
\begin{equation}  \label{constraint}
  \frac{u^{l+1}_{m+M-1}}{u^{l}_{m}} = \frac{w^{l+1}_{m+M}}{w^{l}_{m-1}}, 
\end{equation}
where $m$ and $l$ denote  discrete space and time variables, respectively, $M$ means a range of interaction and $\delta $ is a difference interval.  The time evolution of $u^l_m$ and $w^l_m$ is described by (\ref{nonlineardPPS}), and Eq.(\ref{constraint}) gives a consistent constraint.  We call (\ref{nonlineardPPS}) and (\ref{constraint}) a discrete PPS (dPPS) equation.  It is noted that dPPS equation reduces to the discrete hungry Lotka-Volterra equation when $w^l_m=0$.  
By applying the variable transformations, 
\begin{equation} 
  u^l_m=e^{U^l_m/\varepsilon}, \quad w^l_m=e^{W^l_m/\varepsilon}, \quad \delta=e^{-1/\varepsilon}
\end{equation}
and taking the ultradiscrete limit\cite{Tokihiro}, i.e., $\varepsilon \to +0$, (\ref{nonlineardPPS}) and (\ref{constraint}) reduce to the ultradiscrete PPS (uPPS) equation, 
\begin{equation}  \label{nonlinearuPPS}
\begin{aligned}
  U^{l+1}_{m}-U^{l}_{m}=& \sum_{k=1}^M \left( \max(1, U^{l}_{m-k})- \max(1, U^{l+1}_{m+k}) + \max(1, W^{l+1}_{m+k}) -\max(1, W^{l}_{m-k})\right), \\
  W^{l+1}_{m}-W^{l}_{m}=& \sum_{k=1}^M \left( \max(1, U^{l}_{m-k+1})- \max(1, U^{l+1}_{m+k-1}) + \max(1, W^{l+1}_{m+k-1}) -\max(1, W^{l}_{m-k+1})\right), 
\end{aligned}
\end{equation}
\begin{equation}  \label{uconstraint}
  U^{l+1}_{m+M-1}-U^{l}_{m} = W^{l+1}_{m+M}-W^{l}_{m-1}.  
\end{equation}
The ultradiscrete analogue of PPS solutions for the above system are also constructed.  Moreover, we show that the PPS solutions for uhLV equation are derived from the ones for the above uPPS equation as a special case.

The contents of this article are as follows.  In section 2, we give the discrete DKP equation and its solution.  In section 3, we derive the dPPS equation and solution from the discrete DKP.  In section 4, we ultradiscretize the PPS solution.  In section 5, we show the relation between the obtained solution and that of the uhLV equation.  Finally, the concluding remarks are given in section 6.  
\section{Discrete DKP Equation and Solution}
Let us start from the discrete DKP equation\cite{DKP, dDKP}, which is expressed by 
\begin{equation}  \label{DKP}
\begin{aligned}
  &\delta (\tau (k+1, l+1, m+1)\tau (k, l, m)-\tau (k+1, l+1, m)\tau (k, l, m+1))\\
  =&\tau (k+1, l, m+1)\tau (k, l+1, m)-\tau (k+1, l, m)\tau (k, l+1, m+1),  
\end{aligned}
\end{equation}
where $\delta $ is a constant.  We have the pfaffian solution to discrete DKP equation,  
\begin{equation}  \label{DKP sol1}
  \tau(k, l, m)= \pf(\alpha_1, \alpha_2, \dots , \alpha_N, \beta_N , \dots , \beta_2, \beta_1)_{k, l, m},   
\end{equation}
where the pfaffian elements are defined by
\begin{equation}  \label{DKP sol1 def1}
  \pf(\alpha_i, \alpha_j)_{k, l} = a_{ij} \eta_i(k, l)\eta_j(k, l), \quad   \pf(\alpha_i, \beta_j)_{k, l, m} = \delta_{i, j}+ \eta_i(k, l)\psi_j(m), \quad   \pf(\beta_i, \beta _j)_{m} = b_{ij} \psi_{ij}(m), 
\end{equation}
\begin{equation}  \label{DKP sol1 def2}
\begin{aligned}
  &a_{ij} =\frac{q_j-q_i}{1-q_iq_j},  \qquad   \eta_i(k, l)=q_i^k\omega _i^lc_i, \qquad  \omega _i= \frac{q_i+\delta }{1+\delta q_i}, \\
  &\delta _{i, j} = \begin{cases} 1 & (i=j) \\ 0 & (i\not=j)\end{cases} , \qquad  \psi_i(m) = p_i^m\phi_i(m),  \qquad  b_{ij} =\frac{1}{1-p_i^Mp_j^M}, \\
  &\psi_{ij}(m) = \sum _{1\le \mu\le M} (\psi _i(m+\mu)\psi _j(m+\mu-1)-\psi _i(m+\mu-1)\psi _j(m+\mu)),   
\end{aligned}
\end{equation}
where $p_i$, $q_i$ and $c_i$ are arbitrary constants, $M$ is a positive integer, and  $\phi_i(m)$ is a periodic function which satisfies $\phi_i(m+M)=\phi_i(m)$.  Here and hereafter we denote $k$, $l$ and $m$ as suffix, and may omit unshifted independent variables for simplicity.  Then the pfaffian (\ref{DKP sol1}) satisfies Eq.(\ref{DKP}) (see Appendix B).  
\section{Discrete PPS Equation and Solution}
We shall reduce the discrete DKP equation and solution given in the previous section into the dPPS equation and solution.  
\begin{proposition}  \label{prop1}
Suppose $q_i= p_i^M$ in (\ref{DKP sol1 def2}).  Then the pfaffian (\ref{DKP sol1}) satisfies
\begin{equation}
  \tau (k+1, l, m)=\tau (k, l, m+M).  
\end{equation}
\end{proposition}
\textbf{Proof.} \ Let us use the triangle matrix notation, 
\begin{equation}
\begin{aligned}
  \tau(k, l, m) = &
\left.
\begin{array}{ccccccc}
    | \ \pf(\alpha_1, \alpha_2) & \dots & \pf(\alpha_1, \alpha_N) & \pf(\alpha_1, \beta_N) & \pf(\alpha_1, \beta_{N-1}) & \dots & \pf(\alpha_1, \beta_1) \\
    & \ddots & \vdots  & \vdots & \vdots &  & \vdots  \\
   & & \pf(\alpha_{N-1}, \alpha_N) & \pf(\alpha_{N-1}, \beta_N) & \pf(\alpha_{N-1}, \beta_{N-1})  & \ldots  & \pf(\alpha_{N-1}, \beta_1)  \\
   & &  & \pf(\alpha _N, \beta _N) & \pf(\alpha_N, \beta_{N-1}) & \ldots & \pf(\alpha _N, \beta _1)  \\
   & &  &  &  \pf(\beta _N, \beta _{N-1}) & \ldots & \pf(\beta _N, \beta _1)  \\
   & &  &  &   & \ddots & \vdots  \\
   & &  &  &   &  & \pf(\beta _2, \beta _1)  
\end{array}
\right|\\
  =&
\left.
\begin{array}{cc}
    | \ \pf(\alpha_i, \alpha_j) & \pf(\alpha_i, \beta_j) \\
   & \pf(\beta _i, \beta _j)  
\end{array}
\right|.   
\end{aligned}
\end{equation}
Then we have 
\begin{equation}  \label{tau(k+1)}
\begin{aligned}
  \tau(k+1, l, m)  =&
\left.
\begin{array}{cc}
    | \ \pf(\alpha_i, \alpha_j)_{k+1} & \pf(\alpha_i, \beta_j)_{k+1} \\
   & \pf(\beta _i, \beta _j)_{k+1}  
\end{array}
\right| \\
   =&
\left.
\begin{array}{cc}
    | \ p_i^Mp_j^M \pf (\alpha_i, \alpha_j)_k & \delta _{i, j}+p_i^M\eta_i(k, l)\psi_j(m) \\
   & \pf(\beta _i, \beta _j)_k  
\end{array}
\right|   \\
  =& 
\left( \prod _{i=1}^N p_i^M \right)\left.
\begin{array}{cc}
    | \ \pf(\alpha_i, \alpha_j)_k & \delta _{i, j}/p_i^M+\eta_i(k, l)\psi_j(m) \\
   & \pf(\beta _i, \beta _j)_k  
\end{array}
\right|.   
\end{aligned}
\end{equation}
On the other hand, we also have 
\begin{equation}
\begin{aligned}
  \tau(k, l, m+M)  =&
\left.
\begin{array}{cc}
    | \ \pf(\alpha_i, \alpha_j)_{m+M} & \pf(\alpha_i, \beta_j)_{m+M} \\
   & \pf(\beta _i, \beta _j)_{m+M}  
\end{array}
\right|   \\
  =&
\left.
\begin{array}{cc}
    | \ \pf(\alpha_i, \alpha_j)_m & \delta _{i, j}+p_j^M\eta_i(k, l)\psi_j(m) \\
   & p_i^Mp_j^M \pf(\beta _i, \beta _j)_m  
\end{array}
\right|   \\
  =& 
\left(\prod _{i=1}^N p_i^M\right) \left.
\begin{array}{cc}
    | \ \pf(\alpha_i, \alpha_j)_m & \delta _{i, j}/p_j^M+\eta_i(k, l)\psi_j(m) \\
   & \pf(\beta _i, \beta _j)_m  
\end{array}
\right|,   
\end{aligned}
\end{equation}
which is identical with (\ref{tau(k+1)}).  This completes the proof.  \par 
Proposition \ref{prop1} shows the pfaffian (\ref{DKP sol1}) under the condition $q_i=p_i^M$ also satisfies 
\begin{equation}  \label{DKP2}
\begin{aligned}
  &\delta (\tau (k, l+1, m+1+M)\tau (k, l, m)-\tau (k, l+1, m+M)\tau (k, l, m+1))\\
  =&\tau (k, l, m+1+M)\tau (k, l+1, m)-\tau (k, l, m+M)\tau (k, l+1, m+1).    
\end{aligned}
\end{equation}
Denote $f_m^l$ as $\tau (k, l, m-Ml)$.  Then Eq.(\ref{DKP2}) is reduced into   
\begin{equation}  \label{dPPS}
  \delta  (f_{m+M+1}^{l+1}f_{m-M}^l -  f_{m+M}^{l+1}f_{m-M+1}^l )  = f_{m+1}^l f_{m}^{l+1}  - f_m^lf_{m+1}^{l+1}.  
\end{equation}
Thus we have the following proposition\cite{RIAM}.  
\begin{proposition}\label{prop2}
Let $f^l_m$ be 
\begin{equation}  \label{dPPS sol}
  f^l_m=\pf(a_1, a_2, \dots , a_N, b_N, \dots , b_2, b_1)_{l, m}, 
\end{equation}
where 
\begin{equation}  \label{dPPS sol def1}
\begin{aligned}
  \pf(a_i, a_j) = a_{ij} s_i(l, m)s_j(l, m), \quad  \pf(a_i, b_j) = \delta_{i, j}+ s_i(l, m)\phi_j(m), \quad   \pf(b_i, b_j) =& b_{ij} \phi_{ij}(m),  
\end{aligned}
\end{equation}
\begin{equation}  \label{dPPS sol def2}
\begin{aligned}
&s_i(l, m)=p_i^{m-Ml}\omega _i^lc_i,  \\
&\phi_{ij}(m) = \sum _{1\le \mu\le M} (p_i^{\mu }p_j^{\mu -1}\phi_i(m+\mu)\phi_j(m+\mu-1)-p_i^{\mu -1}p_j^{\mu }\phi_i(m+\mu-1)\phi_j(m+\mu)).   
\end{aligned}
\end{equation}
The definitions of other symbols are the same as (\ref{DKP sol1 def2}) with $q_i=p_i^M$.  Then, the pfaffian (\ref{dPPS sol}) satisfies Eq.(\ref{dPPS}).  
\end{proposition}
\textbf{Proof.} \ We show $\tau(k, l, m-Ml)$ coincides with the pfaffian (\ref{dPPS sol}).  In a similar manner to proposition \ref{prop1}, we have 
\begin{equation}
\begin{aligned}
  \tau(k, l, m-Ml)  =&
\left.
\begin{array}{cc}
    | \ \pf(\alpha_i, \alpha_j)_{m-Ml} & \pf(\alpha_i, \beta_j)_{m-Ml} \\
   & \pf(\beta _i, \beta _j)_{m-Ml}  
\end{array}
\right|   \\
  =&
\left.
\begin{array}{cc}
    | \ a_{ij} p_i^{kM}\omega _i^lc_ip_j^{kM}\omega _j^lc_j & \delta _{i, j}+p_i^{kM}\omega _i^lc_i p_j^{m-Ml}\phi_j(m) \\
   & p_i^{m-Ml}p_j^{m-Ml} b_{ij}\phi_{ij}  
\end{array}
\right|.   
\end{aligned}
\end{equation}
Replacing $c_i$ as $p_i^{-kM} c_i$, it is rewritten as  
\begin{equation}  \label{tau(m-Ml)}
\begin{aligned}
 &  \left.
\begin{array}{cc}
    | \ a_{ij} \omega _i^lc_i\omega _j^lc_j & \delta _{i, j}+\omega _i^lc_i p_j^{m-Ml}\phi_j(m) \\
   & p_i^{m-Ml}p_j^{m-Ml} b_{ij}\phi_{ij}  
\end{array}
\right|   \\
  =& \left( \prod _{i=1}^N p_i^{m-Ml}\right)
\left.
\begin{array}{cc}
    | \ a_{ij} \omega _i^lc_i\omega _j^lc_j & \delta _{i, j}/p_j^{m-Ml}+\omega _i^lc_i \phi_j(m) \\
   & b_{ij} \phi_{ij}  
\end{array}
\right|.  
\end{aligned}
\end{equation}
On the other hand, we also have
\begin{equation}
\begin{aligned}
    f^l_m  =&  
\left.
\begin{array}{cc}
    | \ \pf(a_i, a_j) & \pf(a_i, b_j) \\
   & \pf(b_i, b_j)  
\end{array}
\right|   \\
  =&\left.
\begin{array}{cc}
    | \ a_{ij} p_i^{m-Ml}\omega _i^lc_ip_j^{m-Ml}\omega _j^lc_j & \delta _{i, j}+p_i^{m-Ml}\omega _i^lc_i \phi_j(m) \\
   & b_{ij}\phi_{ij}  
\end{array}
\right|   \\
  =& \left( \prod _{i=1}^N p_i^{m-Ml}\right)
\left.
\begin{array}{cc}
    | \ a_{ij} \omega _i^lc_i\omega _j^lc_j & \delta _{i, j}/p_i^{m-Ml}+\omega _i^lc_i \phi_j(m) \\
   & b_{ij} \phi_{ij}  
\end{array}
\right|,   
\end{aligned}
\end{equation}
which is identical with (\ref{tau(m-Ml)}).  This completes the proof.  \par 
Now we have the following proposition.  
\begin{proposition}\label{prop1.2}
Eq.(\ref{dPPS}) derives (\ref{nonlineardPPS}) and (\ref{constraint}) through transformations  
\begin{equation}  \label{transformation}
  u^l_m = \frac{f^l_{m-M}f^{l+1}_{m+M+1}}{f^l_{m}f^{l+1}_{m+1}}, \qquad w^l_m = \frac{f^{l+1}_{m+M}f^{l}_{m-M+1}}{f^l_{m+1}f^{l+1}_{m}}.   
\end{equation}
\end{proposition}
\textbf{Proof.} \ Eq.(\ref{constraint}) follows from (\ref{transformation}) straightforwardly.  Eq.(\ref{dPPS}) is rewritten as 
\begin{equation}
\begin{aligned}
  1+\delta u^l_m =& \frac{f^l_{m+1}f^{l+1}_m}{f^l_{m}f^{l+1}_{m+1}} +\delta \frac{f^{l+1}_{m+M}f^{l}_{m-M+1}}{f^l_{m}f^{l+1}_{m+1}},  \\
  1+\delta w^l_m =& \frac{f^l_{m}f^{l+1}_{m+1}}{f^l_{m+1}f^{l+1}_{m}} +\delta \frac{f^{l+1}_{m+M+1}f^{l}_{m-M}}{f^l_{m+1}f^{l+1}_{m}}.   
\end{aligned}
\end{equation}
In particular, 
\begin{equation}
  \frac{1+\delta u^l_m}{1+\delta w^l_m} = \frac{f^l_{m+1}f^{l+1}_m}{f^l_{m}f^{l+1}_{m+1}}
\end{equation}
holds.  Thus, we have
\begin{equation}
  \prod _{k=1}^M \frac{1+\delta u^{l}_{m-k}}{1+\delta u^{l+1}_{m+k}}\frac{1+\delta w^{l+1}_{m+k}}{1+\delta w^{l}_{m-k}} = \prod _{k=1}^M \frac{f^l_{m+1-k}f^{l+1}_{m-k}}{f^l_{m-k}f^{l+1}_{m+1-k}} \frac{f^{l+1}_{m+k}f^{l+2}_{m+1+k}}{f^{l+1}_{m+1+k}f^{l+2}_{m+k}} = \frac{u^{l+1}_m}{u^l_m}.     
\end{equation}
The other equation can be obtained similarly.  This completes the proof.  \par
\begin{figure}  
  \begin{center}
   \begin{picture}(400, 120)
  \put(0,10){\includegraphics[width=4cm, height=3cm, clip]{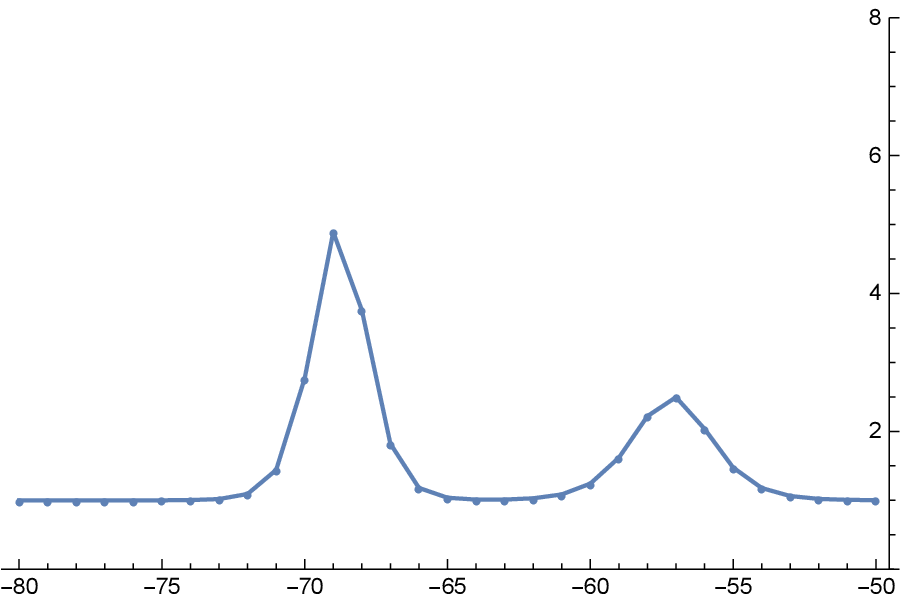}} 
  \put(150,10){\includegraphics[width=4cm, height=3cm, clip]{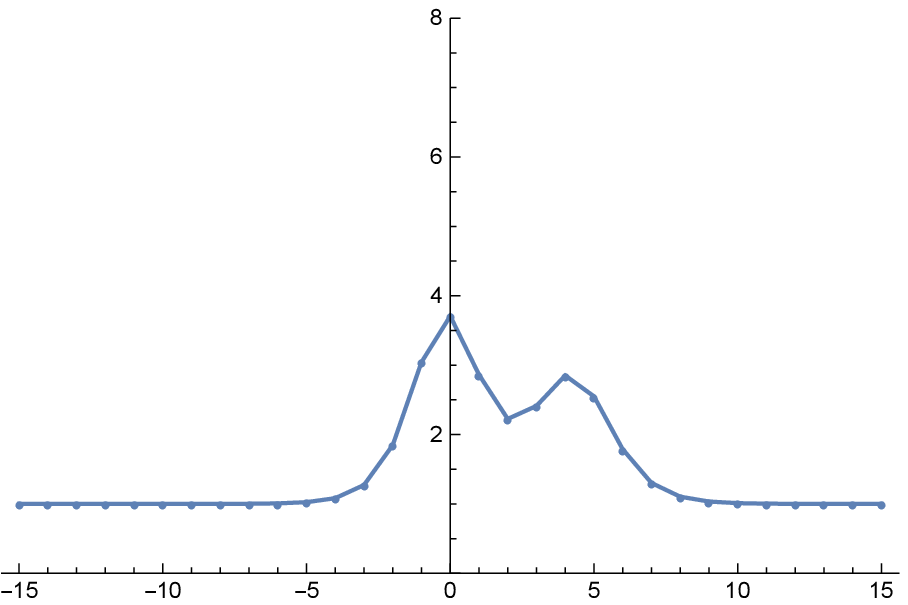}} 
  \put(300,10){\includegraphics[width=4cm, height=3cm, clip]{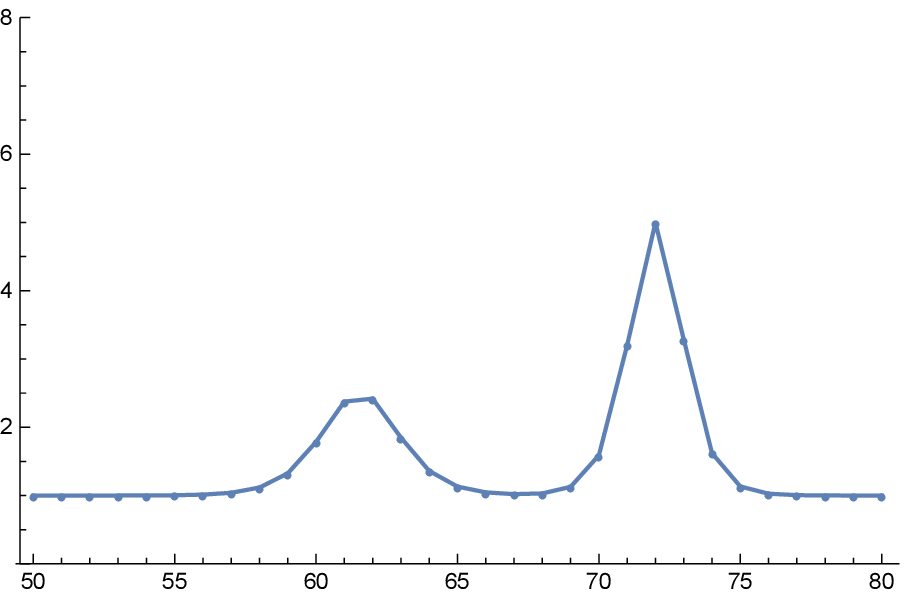}} 
  \put(35,0){\scriptsize $l=-50$} 
  \put(200,0){\scriptsize $l=0$} 
  \put(350,0){\scriptsize $l=50$} 
   \end{picture}
\caption{$2$-soliton solution $u^l_m$.   }
\label{fig11}
  \end{center}
\end{figure}

\begin{figure}  
  \begin{center}
   \begin{picture}(400, 120)
  \put(0,10){\includegraphics[width=4cm, height=3cm, clip]{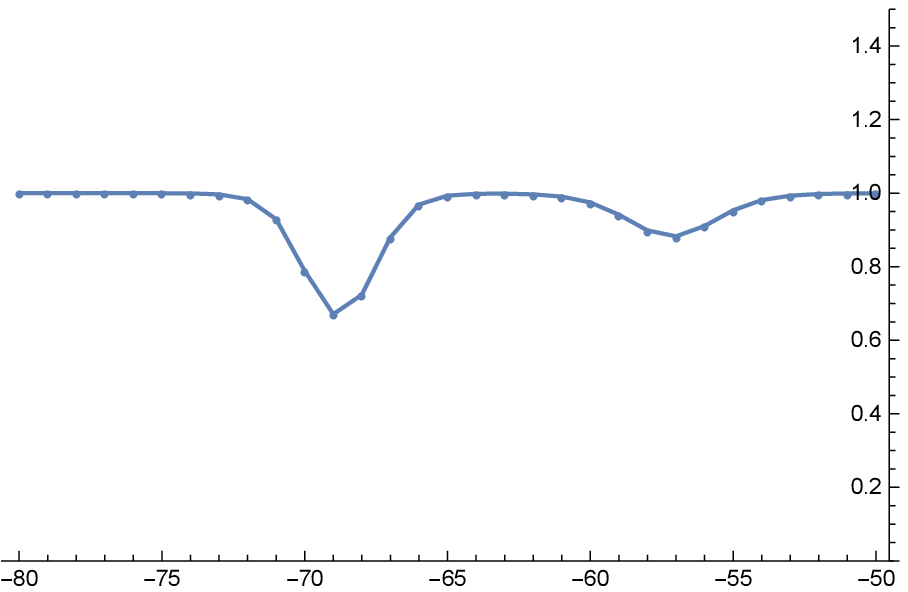}} 
  \put(150,10){\includegraphics[width=4cm, height=3cm, clip]{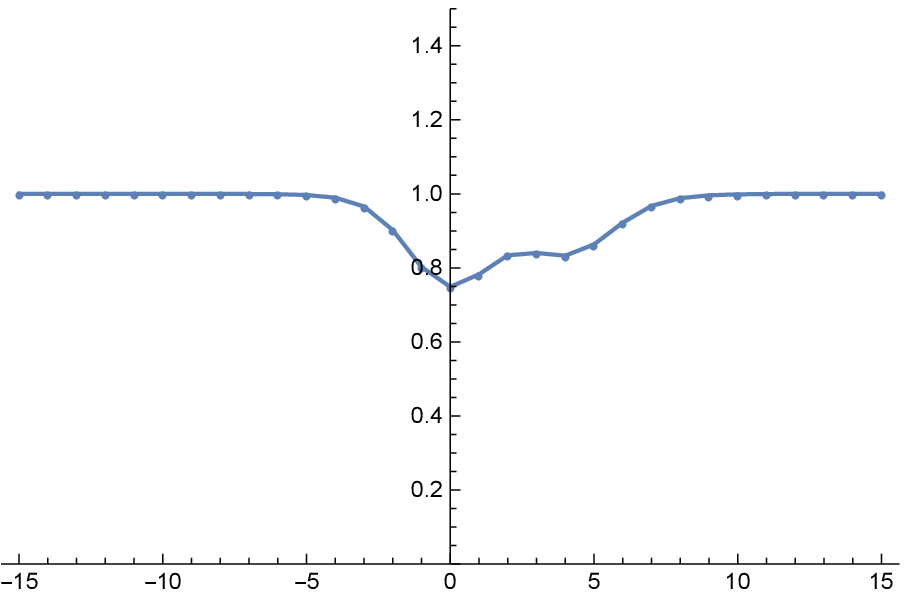}} 
  \put(300,10){\includegraphics[width=4cm, height=3cm, clip]{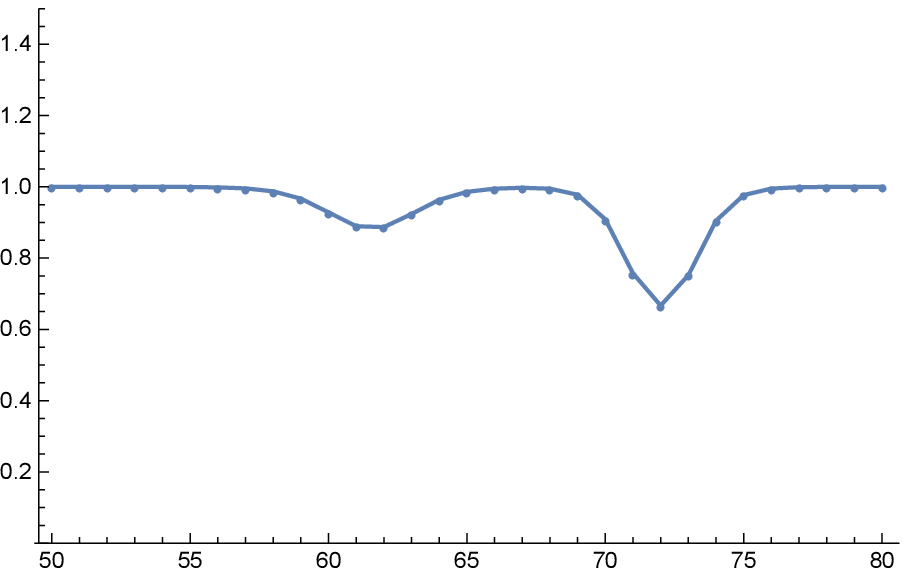}} 
  \put(35,0){\scriptsize $l=-50$} 
  \put(200,0){\scriptsize $l=0$} 
  \put(350,0){\scriptsize $l=50$} 
   \end{picture}
\caption{$2$-soliton solution $w^l_m$.   }
\label{fig12}
  \end{center}
\end{figure}

\begin{figure}  
  \begin{center}
   \begin{picture}(400, 120)
  \put(0,10){\includegraphics[width=4cm, height=3cm, clip]{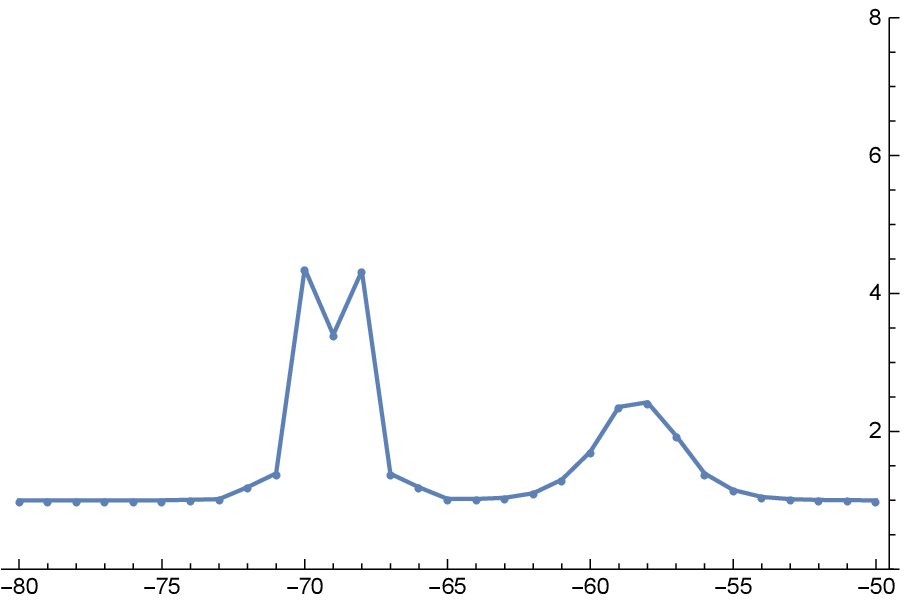}} 
  \put(150,10){\includegraphics[width=4cm, height=3cm, clip]{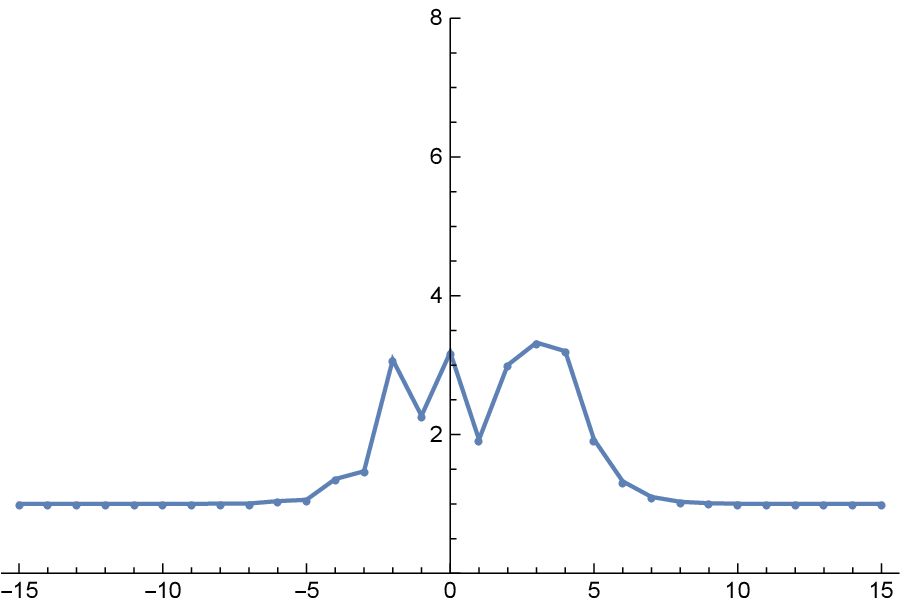}} 
  \put(300,10){\includegraphics[width=4cm, height=3cm, clip]{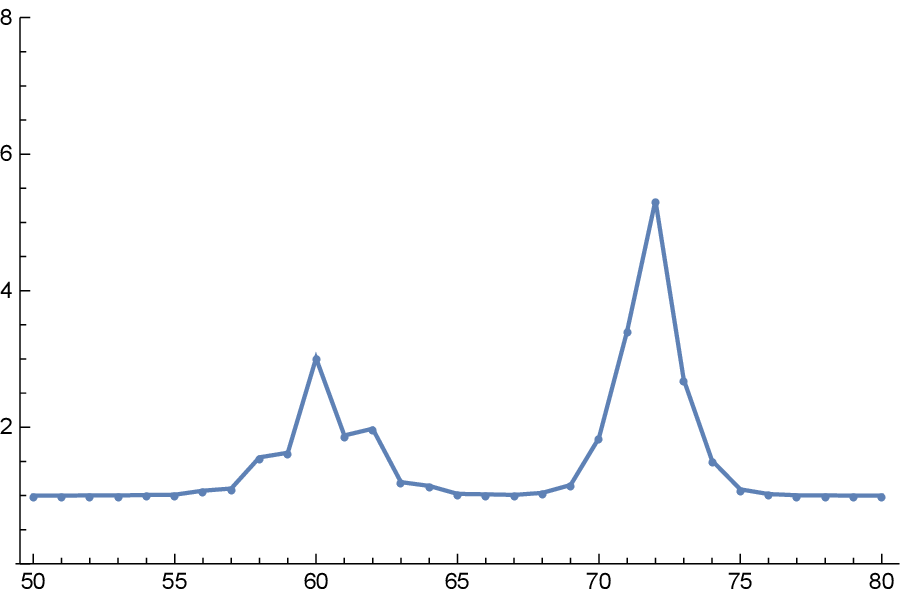}} 
  \put(35,0){\scriptsize $l=-50$} 
  \put(200,0){\scriptsize $l=0$} 
  \put(350,0){\scriptsize $l=50$} 
   \end{picture}
\caption{$2$-dPPS solution $u^l_m$.   }
\label{fig21}
  \end{center}
\end{figure}

\begin{figure}  
  \begin{center}
   \begin{picture}(400, 120)
  \put(0,10){\includegraphics[width=4cm, height=3cm, clip]{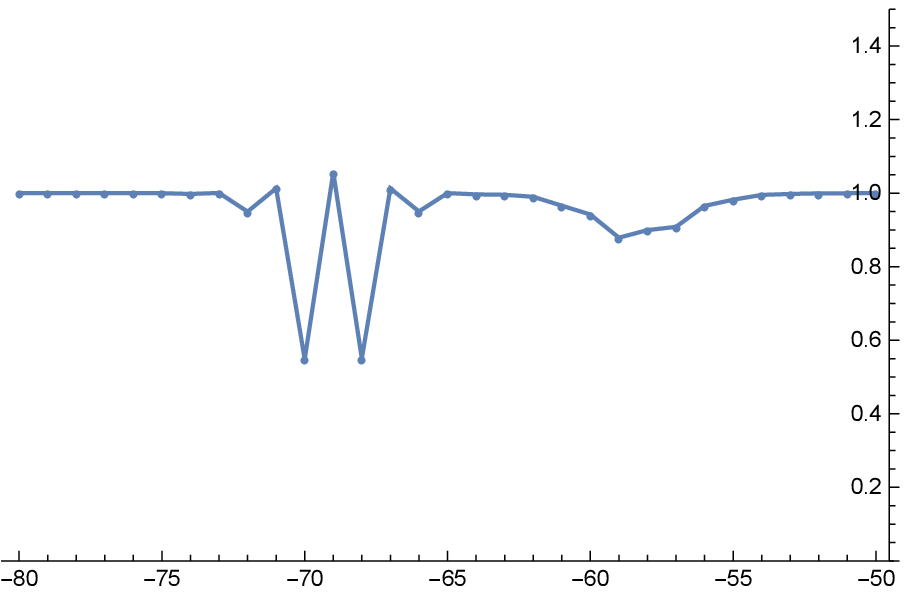}} 
  \put(150,10){\includegraphics[width=4cm, height=3cm, clip]{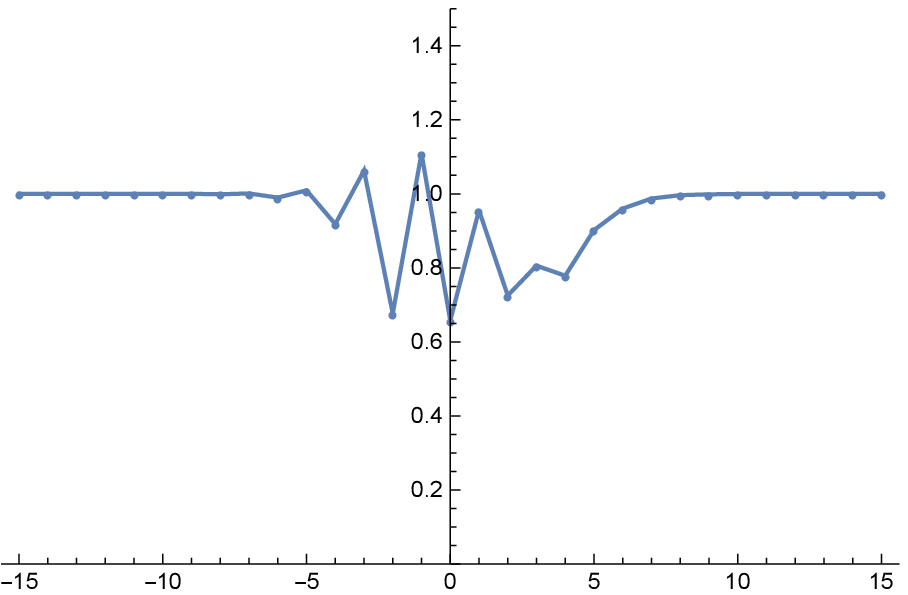}} 
  \put(300,10){\includegraphics[width=4cm, height=3cm, clip]{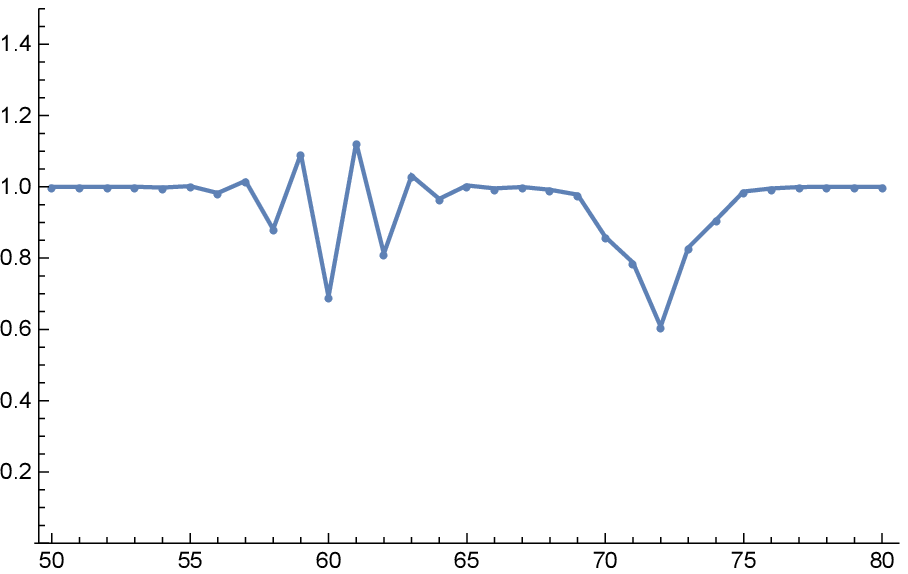}} 
  \put(35,0){\scriptsize $l=-50$} 
  \put(200,0){\scriptsize $l=0$} 
  \put(350,0){\scriptsize $l=50$} 
   \end{picture}
\caption{$2$-dPPS solution $w^l_m$.   }
\label{fig22}
  \end{center}
\end{figure}

  
Due to proposition \ref{prop1.2}, we may call (\ref{dPPS}) as the dPPS equation in bilinear form. Thus we have derived the pfaffian solution to the dPPS equation.  Figures \ref{fig11}, \ref{fig12}, \ref{fig21} and \ref{fig22} show the behaviour of the dPPS solution $u^l_m$ and $w^l_m$ in the case of $N=M=2$.  Here $l$ and $m$ denote time and space variables respectively.  In Fig.\ref{fig11} and Fig.\ref{fig12}, we set $\delta =1/3$, $(p_1, p_2, c_1, c_2)=(5, 3, 1, 1)$ and $\phi_i(m)=1$ for $i=1, 2$.  They behave as the usual soliton solutions.  In Fig.\ref{fig21} and Fig.\ref{fig22},  we set $\delta =1/3$, $(p_1, p_2, c_1, c_2)=(5, 3, 1, 1)$ and 
\begin{equation}
\phi_1(m) =
\begin{cases}
  1 & (\text{$m$: even})\\
  2 & (\text{$m$: odd})
\end{cases}, \qquad 
\phi_2(m) =
\begin{cases}
  3 & (\text{$m$: even})\\
  5 & (\text{$m$: odd})
\end{cases}.
\end{equation}
We can observe that the high jaggy wave collides with low one and overtakes.  The shape of the solitary wave changes periodically.  
\section{Ultradiscretization}
In this section we ultradiscretize the pfaffian solution (\ref{dPPS sol}).  For ultradiscretizing we assume $0<p_1< p_2<\dots < p_N<1$, $0<c_i$, $0<\phi_i(m)$, $0<\delta$ and 
\begin{equation}  \label{cond phi0}
\begin{aligned}
  \frac{p_i}{p_j} >&  \max_{1\le m\le M}\frac{\phi_i(m -1)\phi_j(m)}{\phi_i(m)\phi_j(m-1)} , \\
  p_ip_j\le & \min_{1\le m\le M}\frac{\phi_i(m) \phi_j(m -1)}{\phi_i(m+1)\phi_j(m )},
\end{aligned}
\end{equation}
for $i>j$.  In the case of usual soliton, i.e., when $\phi_i(m)$ is independent of $m$, the above condition (\ref{cond phi0}) is obviously satisfied.  From (\ref{cond phi0}), we have 
\begin{equation}  \label{cond phi}
\begin{aligned}
  &p_i\phi_i(m)\phi_j(m-1) > p_j\phi_i(m -1)\phi_j(m), \\
  &\phi_i(m) \phi_j(m -1) \ge p_ip_j\phi_i(m+1)\phi_j(m ),
\end{aligned}
\end{equation}
for $i>j$ and $m=1, 2, \dots , M$.  Now we introduce reparametrizations, 
\begin{equation}  \label{ud trans}
  p_i=e^{-P_i/\varepsilon }, \quad \delta=e^{-1/\varepsilon }, \quad \omega_i=e^{\Omega_i/\varepsilon }, \quad c_i=e^{C_i/\varepsilon }, \quad \phi_i(m)=e^{\Phi_i(m)/\varepsilon },  
\end{equation}
and we are going to take a limit $\varepsilon \to +0$ for ultradiscretization.  It is commented that $P_i$ is positive, $\Omega_i$ is negative and $P_i<P_j$ for $i>j$.  In the following, we show the ultradiscrete limit of PPS solutions, first for $N=2$, $3$ and next for general $N$.  
\subsection{Case of $N=2$}
\begin{figure}  
  \begin{center}
   \begin{picture}(400, 110)
  \put(0,10){\includegraphics[width=4cm, height=3cm, clip]{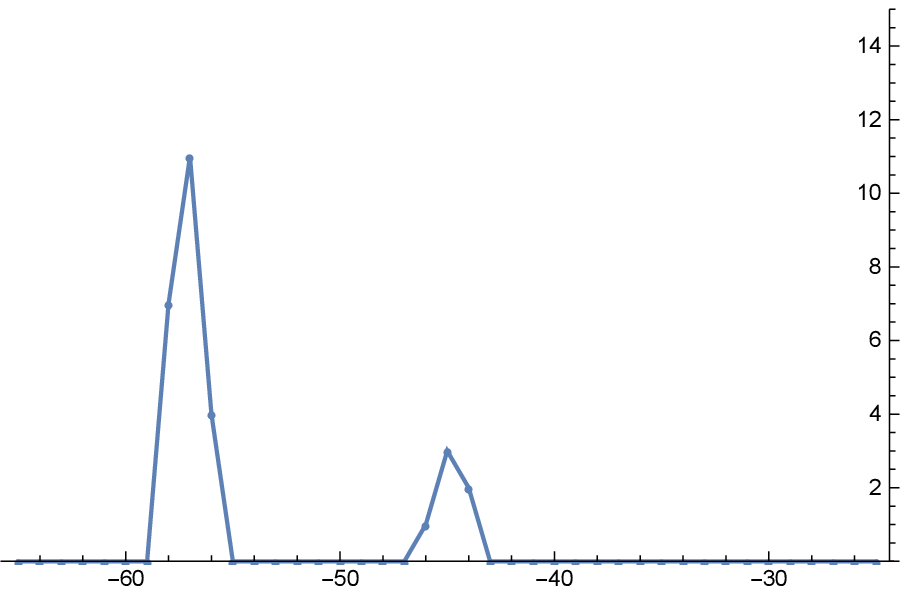}} 
  \put(150,10){\includegraphics[width=4cm, height=3cm, clip]{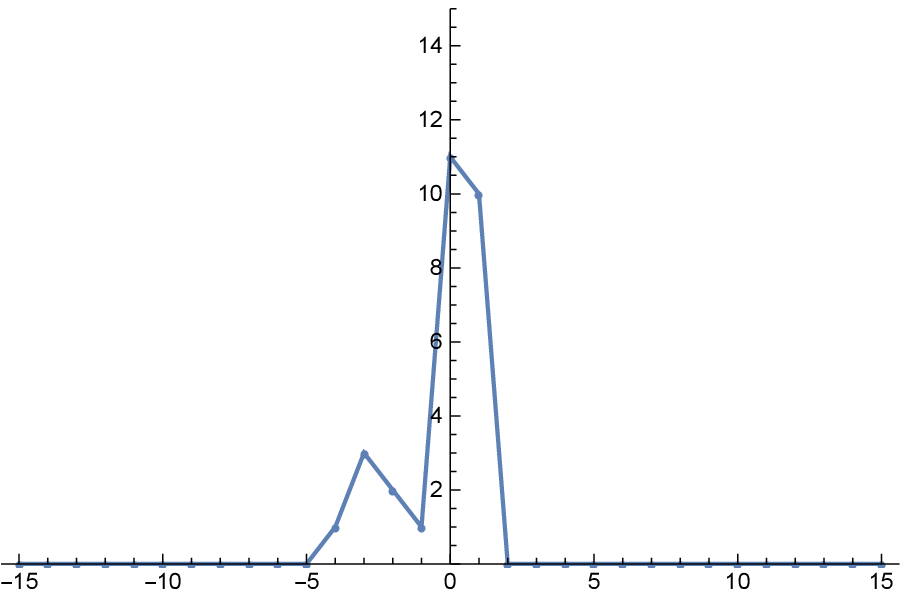}} 
  \put(300,10){\includegraphics[width=4cm, height=3cm, clip]{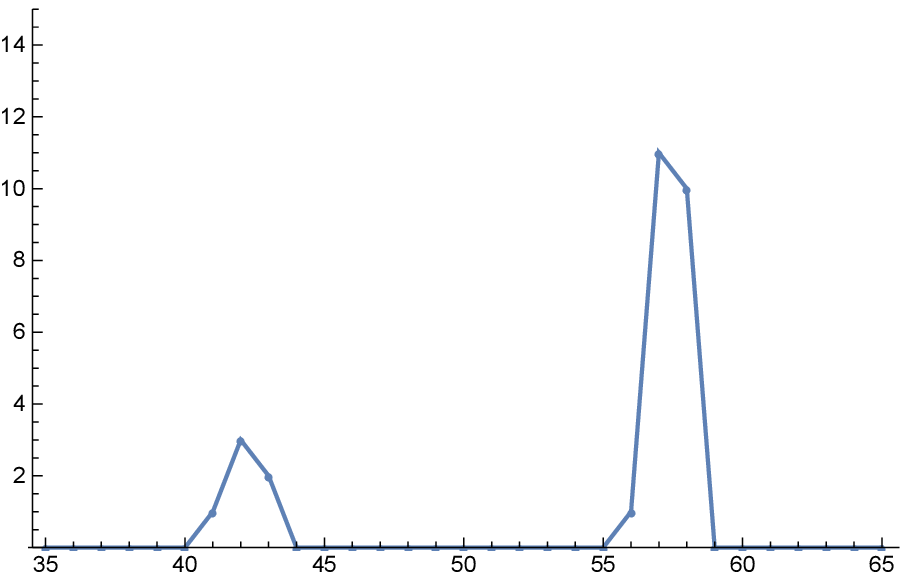}} 
  \put(35,0){\scriptsize $l=-35$} 
  \put(200,0){\scriptsize $l=-5$} 
  \put(350,0){\scriptsize $l=25$} 
   \end{picture}
\caption{$2$-soliton solution $U^l_m$.      }
\label{fig31}
  \end{center}
\end{figure}

\begin{figure}  
  \begin{center}
   \begin{picture}(400, 110)
  \put(0,10){\includegraphics[width=4cm, height=3cm, clip]{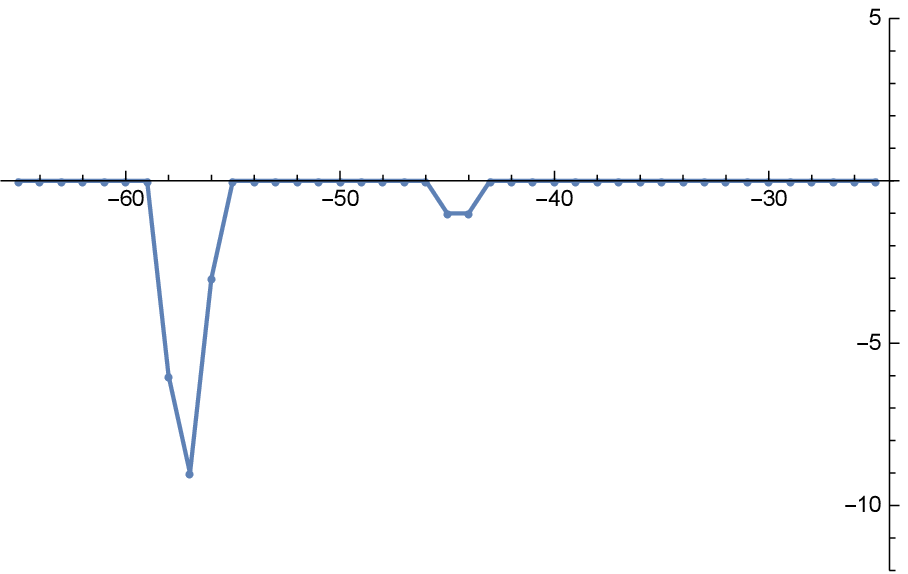}} 
  \put(150,10){\includegraphics[width=4cm, height=3cm, clip]{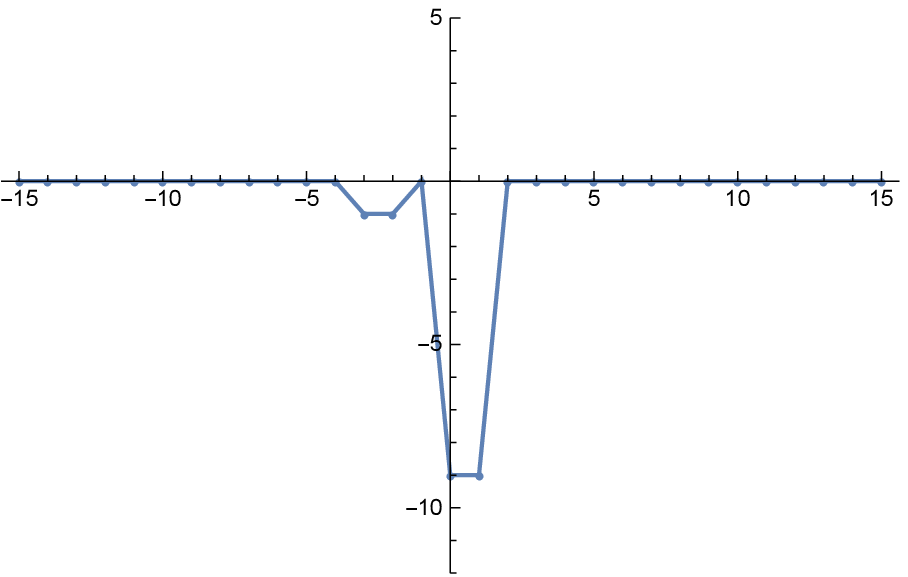}} 
  \put(300,10){\includegraphics[width=4cm, height=3cm, clip]{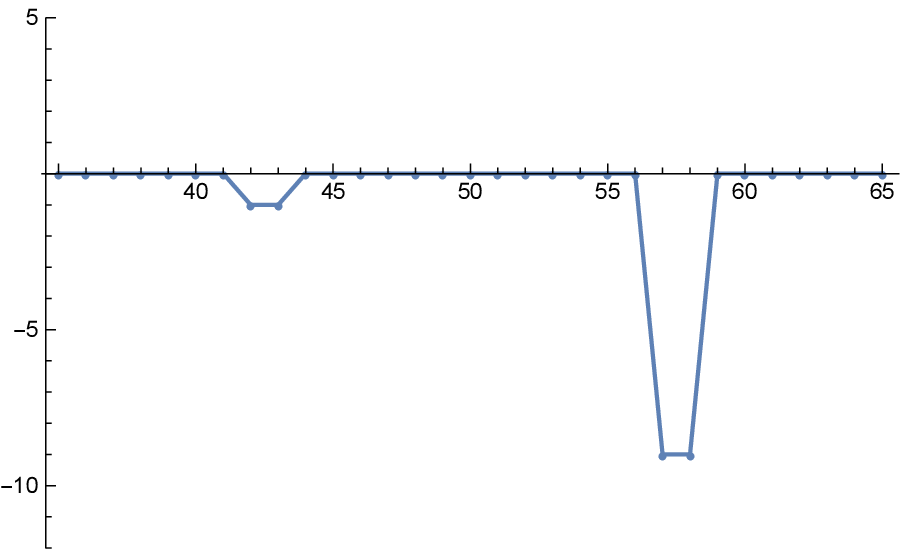}} 
  \put(35,0){\scriptsize $l=-35$} 
  \put(200,0){\scriptsize $l=-5$} 
  \put(350,0){\scriptsize $l=25$} 
   \end{picture}
\caption{$2$-soliton solution $W^l_m$.      }
\label{fig32}
  \end{center}
\end{figure}

\begin{figure}  
  \begin{center}
   \begin{picture}(400, 110)
  \put(0,10){\includegraphics[width=4cm, height=3cm, clip]{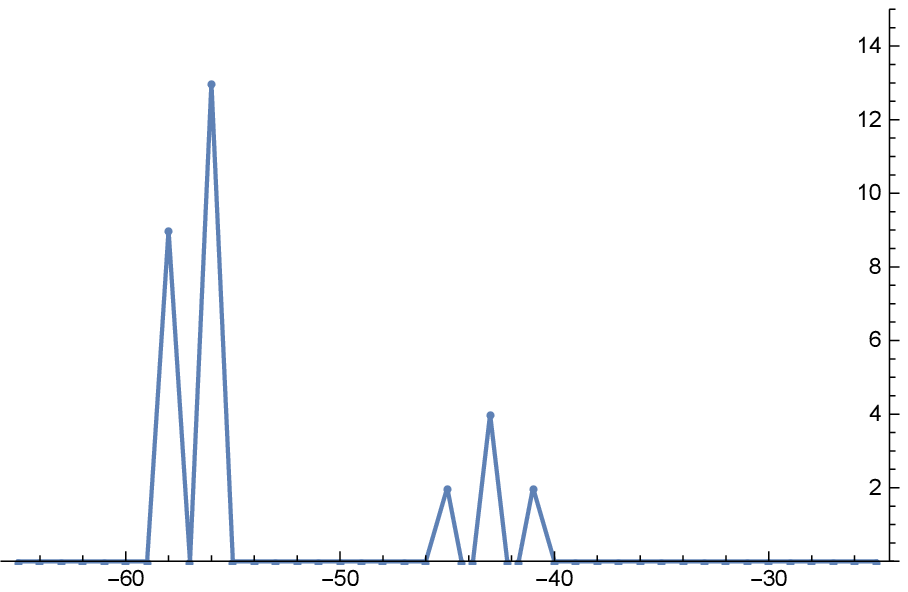}} 
  \put(150,10){\includegraphics[width=4cm, height=3cm, clip]{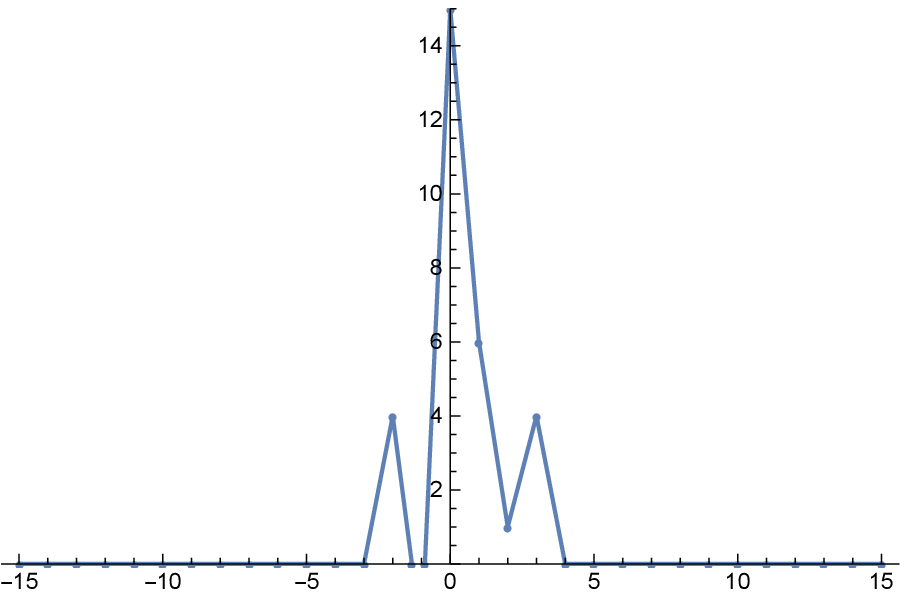}} 
  \put(300,10){\includegraphics[width=4cm, height=3cm, clip]{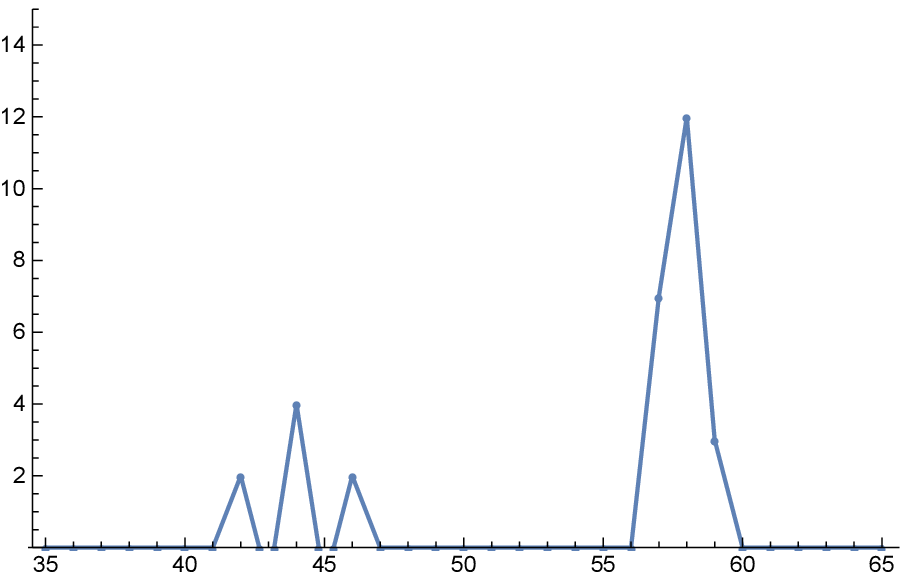}} 
  \put(35,0){\scriptsize $l=-30$} 
  \put(200,0){\scriptsize $l=0$} 
  \put(350,0){\scriptsize $l=30$} 
   \end{picture}
\caption{$2$-uPPS solution $U^l_m$.     }
\label{fig41}
  \end{center}
\end{figure}

\begin{figure}  
  \begin{center}
   \begin{picture}(400, 110)
  \put(0,10){\includegraphics[width=4cm, height=3cm, clip]{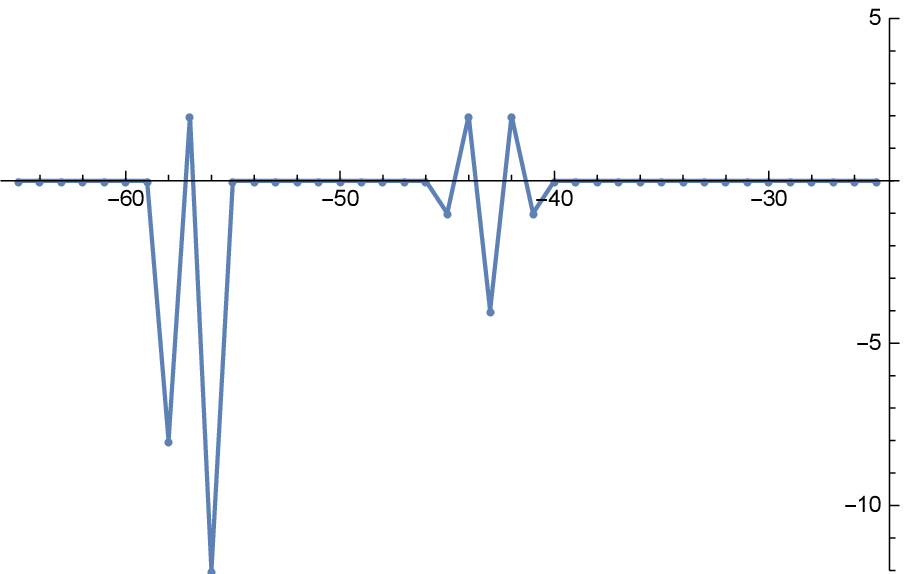}} 
  \put(150,10){\includegraphics[width=4cm, height=3cm, clip]{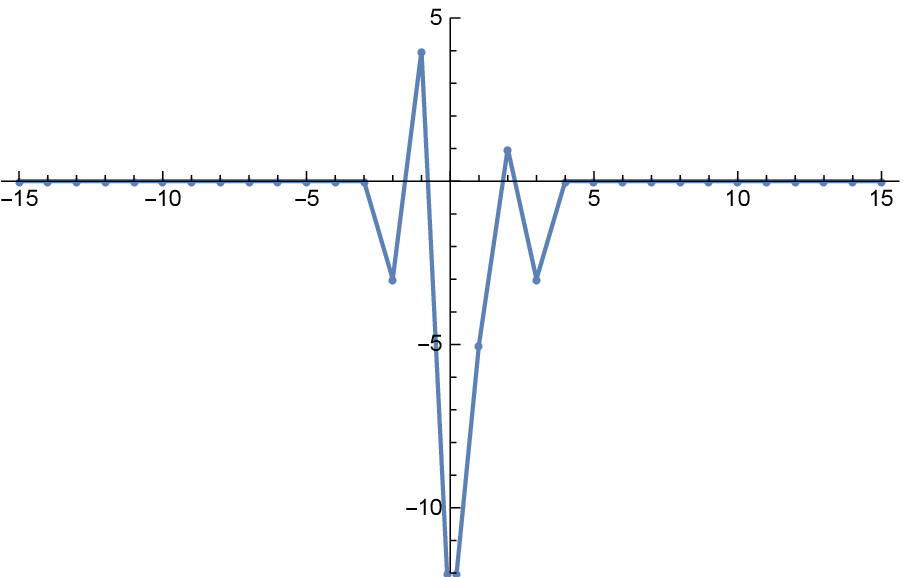}} 
  \put(300,10){\includegraphics[width=4cm, height=3cm, clip]{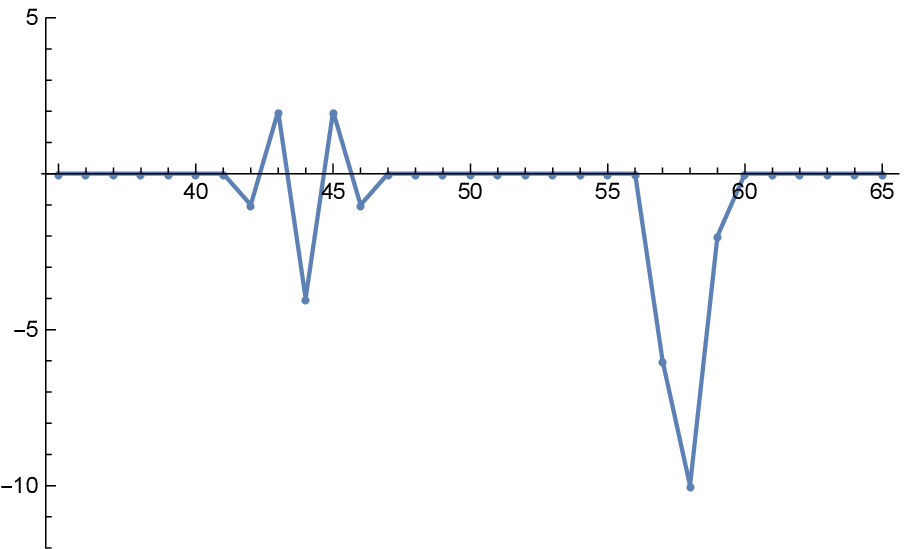}} 
  \put(35,0){\scriptsize $l=-30$} 
  \put(200,0){\scriptsize $l=0$} 
  \put(350,0){\scriptsize $l=30$} 
   \end{picture}
\caption{$2$-uPPS solution $W^l_m$.     }
\label{fig42}
  \end{center}
\end{figure}

In the case of $N=2$, $f^l_m$ is expressed by 
\begin{equation}  \label{2sol}
\begin{aligned}
  f^l_m =&\pf(a_1, a_2, b_2, b_1) \\
  =&1+ s_1\phi_1(m)+s_2\phi_2(m)+a_{12}s_1s_2b_{21}\phi_{21}(m).  
\end{aligned}
\end{equation}
We can ultradiscretize each variable as  
\begin{equation}  
\begin{aligned}
  \omega_i \quad \to \quad &\Omega _i =\max(-MP_i, -1),\\
 s_i(l, m)\quad\to \quad&S_i(l, m)= -(m-Ml)P_i+l\Omega_i+C_i,\\
  a_{12}\quad\to \quad& -MP_2,\\
  b_{21}\quad \to \quad & 0
\end{aligned}
\end{equation}
from (\ref{dPPS sol def2}).  Let us ultradiscretize $\phi_{21}(m)$, which consists of $2M$ terms.  The greatest term among these terms is $p_2\phi_2(m+1)\phi_1(m)$ since  we can derive
\begin{equation}  
\begin{aligned}
  &p_i^{\mu }p_j^{\mu-1}\phi_i(m+\mu)\phi_j(m+\mu-1) > p_i^{\mu-1}p_j^{\mu}\phi_i(m +\mu-1)\phi_j(m+\mu), \\
  &p_i^{\mu}p_j^{\mu-1}\phi_i(m+\mu) \phi_j(m+\mu -1) \ge p_i^{\mu+1}p_j^{\mu}\phi_i(m+\mu+1)\phi_j(m+\mu ), 
\end{aligned}
\end{equation}
for $\mu= 1, 2, \dots , M$ from (\ref{cond phi}).  Thus we obtain    
\begin{equation}
  \phi_{21}(m) \quad \to \quad  -P_2+\Phi_2(m+1)+\Phi_1(m).  
\end{equation}
Therefore the ultradiscrete analogue of (\ref{2sol})  is given by   
\begin{equation}
  f^l_m \to  F^l_m= \max( 0, S_1+\Phi_1(m), S_2+\Phi_2(m), S_1+S_2-(M+1)P_2+\Phi_1(m)+\Phi_2(m+1) ).   
\end{equation}
Note $F^l_m$ leads solutions to the uPPS equation (\ref{nonlinearuPPS}) and (\ref{uconstraint}) through the transformations 
\begin{equation}
  U^l_m = F^l_{m-M}+F^{l+1}_{m+M+1}-F^l_{m}-F^{l+1}_{m+1}, \qquad W^l_m = F^{l+1}_{m+M}+F^{l}_{m-M+1}-F^l_{m+1}-F^{l+1}_{m}.   
\end{equation}
Figures \ref{fig31}, \ref{fig32}, \ref{fig41} and \ref{fig42} show behaviour of the uPPS solution $U^l_m$ and $W^l_m$ in the case of $N=M=2$.  In Fig.\ref{fig31} and Fig.\ref{fig32}, we set $(P_1, P_2, C_1, C_2)=(10, 2, 0, 0)$ and $\Phi_i(m)=0$ for $i=1, 2$.  In Fig.\ref{fig41} and Fig.\ref{fig42}, we set $(P_1, P_2, C_1, C_2)=(10, 2, 0, 0)$ and 
\begin{equation}
\Phi_1(m) =
\begin{cases}
  4 & (\text{$m$: even})\\
  3 & (\text{$m$: odd})
\end{cases}, \qquad 
\Phi_2(m) =
\begin{cases}
  1 & (\text{$m$: even})\\
  6 & (\text{$m$: odd})
\end{cases}.
\end{equation}
They behave similarly to the discrete ones.  
\subsection{Case of $N=3$}
In the case of $N=3$, $f^l_m$ is expanded as
\begin{equation}  \label{3sol}
\begin{aligned}
  f^l_m =&\pf(a_1, a_2, a_3, b_3, b_2, b_1) \\
  =&1+ s_1\phi_1(m)+s_2\phi_2(m)+ s_3\phi_3(m) \\
  & +a_{12}s_1s_2b_{21}\phi_{21}(m)+a_{13}s_1s_3b_{31}\phi_{31}(m)+a_{23}s_2s_3b_{32}\phi_{32}(m)\\
  & +s_1s_2s_3(a_{12}-a_{13}+a_{23})(b_{21}\phi_3(m)\phi_{21}(m)-b_{31}\phi_2(m)\phi_{31}(m)+b_{32}\phi_1(m)\phi_{32}(m)).   
\end{aligned}
\end{equation}
One can prove that $a_{12}-a_{13}+a_{23}$ is equal to $a_{12}a_{13}a_{23}$ and it is ultradiscretized as $-MP_2-2MP_3$.  Moreover, we can determine the greatest term among the expansion of $b_{21}\phi_3\phi_{21}-b_{31}\phi_2\phi_{31}+b_{32}\phi_1\phi_{32}$ as $p_2p_3^2 \phi_1(m)\phi_2(m+1)\phi_3(m+2)$.  It is ultradiscretized as $-P_2-2P_3+\Phi_1(m)+\Phi_2(m+1)+\Phi_3(m+2)$ (see appendix D).  Thus the ultradiscrete analogue of (\ref{3sol}) is obtained by  
\begin{equation}
\begin{aligned}
  &F^l_m =\max\Bigl( 0, S_1+\Phi_1(m), S_2+\Phi_2(m), S_3+\Phi_3(m), \\
  &S_1+S_2-(M+1)P_2+\Phi_1(m)+\Phi_2(m+1), \\
  &S_1+S_3-(M+1)P_3+\Phi_1(m)+\Phi_3(m+1), \\
  &S_2+S_3-(M+1)P_3+\Phi_2(m)+\Phi_3(m+1), \\
  &S_1+S_2+S_3-(M+1)P_2-2(M+1)P_3+\Phi_1(m)+\Phi_2(m+1)+\Phi_3(m+2)\Bigr).  
\end{aligned}
\end{equation}
\subsection{Case of general $N$}
Let us ultradiscretize $f^l_m$ for general $N$.  First, we use the following relation, 
\begin{equation}  \label{rel pf}
  \pf(a_1, \dots , a_N, b_N, \dots , b_1)= \sum _{0\le n\le N} \sum _{I_n \subset \{ 1, 2, \dots , N\} } \pf(a_{i_1}, a_{i_2}, \dots , a_{i_n})\pf(b_{i_n}, \dots , b_{i_2}, b_{i_1}) ,
\end{equation}
where $\sum _{I_n \subset \{ 1, 2, \dots , N\} }$ denotes the summation over all $n$-element subsets $I_n=\{ i_1, i_2, \dots , i_n\}$ chosen from $\{1, 2, \dots , N\}$.  When $n=0$, we define $\sum _{I_n\subset \emptyset }$ is $1$.  When $n$ is odd, we define  
\begin{equation}
  \pf(a_{i_1}, a_{i_2}, \dots , a_{i_n}) = \pf(a_{i_1}, a_{i_2}, \dots , a_{i_n}, \emptyset ),  \quad \pf(b_{i_n}, \dots , b_{i_2}, b_{i_1}) = \pf(\emptyset, b_{i_n}, \dots , b_{i_2}, b_{i_1})
\end{equation} 
with the new entries
\begin{equation}
  \pf(a_i, \emptyset)=s_i(l, m), \qquad \pf(\emptyset , b_i) = \phi_i(m).
\end{equation} 
For example, $f^l_m$ in the case of $N=2$, $3$ are expressed by
\begin{equation}
\begin{aligned}
f^l_m  =&1+\pf(a_1, \emptyset )\pf(\emptyset , b_1)+\pf(a_2, \emptyset )\pf(\emptyset , b_2)+\pf(a_1, a_2 )\pf(b_2 , b_1),\\
 f^l_m =&1+\pf(a_1, \emptyset )\pf(\emptyset , b_1)+\pf(a_2, \emptyset )\pf(\emptyset , b_2)+\pf(a_3, \emptyset )\pf(\emptyset , b_3)\\
  & +\pf(a_1, a_2 )\pf(b_2 , b_1)+\pf(a_1, a_3 )\pf(b_3 , b_1)+\pf(a_2, a_3 )\pf(b_3 , b_2)\\ 
  & +\pf(a_1, a_2, a_3, \emptyset)\pf(\emptyset, b_3, b_2, b_1),
\end{aligned}
\end{equation}
and they correspond to (\ref{2sol}) and (\ref{3sol}).  The proof of the relation (\ref{rel pf}) is given in appendix C.  \par
  Let us consider the expansion of $\pf(a_1, a_2, \dots , a_N)$ and $\pf(b_N, \dots , b_2, b_1)$ respectively.  We assume $N$ is even for simplicity.  Then $\pf(a_1, a_2, \dots , a_N)$ can be expanded as follows\cite{Tsujimoto}.  
\begin{equation}
  \pf(a_1, a_2, \dots , a_N)=\prod _{1\le i<j\le N}a_{i j}\prod _{1\le i\le N}s_i  .
\end{equation}
Hence introducing transformations (\ref{ud trans}) under the assumption $0<p_1< p_2< \dots <p_N<1$, we ultradiscretize  
\begin{equation}
\begin{aligned}
  \pf(a_1, a_2, \dots , a_N) \to  \sum _{1\le i\le N} S_i -\sum _{1\le i<j\le N} MP_j=\sum _{1\le i\le N} S_i -M\sum _{2\le j\le N} (j-1)P_j.  
\end{aligned}
\end{equation}
The expansion of $\pf(b_N, \dots , b_2, b_1)$ cannot be factorized as $\pf(a_1, a_2 , \dots , a_N)$.  However, the greatest term of $\pf(b_N, \dots , b_2, b_1)$ is determined as $\prod _{i=1}^Np_i^{i-1} \phi_i(m+i-1)$ under the assumption (\ref{cond phi}) (see appendix D), thus, 
\begin{equation}
  \pf(b_N, \dots , b_2, b_1) \to \sum _{1\le i\le N}\Bigl( -(i-1)P_i +\Phi_i(m+i-1)\Bigr).
\end{equation}
Therefore we obtain the ultradiscrete analogue of (\ref{rel pf}) as 
\begin{equation}  \label{Nsol}
  F_m^l= \max_{\mu_i=0, 1}\Bigl( \sum_{1\le i\le N}\mu_iS_i(l, m)-(M+1)\sum_{1\le i<j\le N}\mu_i\mu_j P_j+ \sum_{1\le j\le N}\mu_j\Phi_j(m+\sum_{1\le i\le j-1}\mu_i) \Bigr),  
\end{equation}  
where $\max_{\mu_i=0, 1}X(\mu_1, \mu_2, \dots , \mu_N)$ denotes the maximum value of $X$ in $2^N$ possible cases $\{ \mu_1, \mu_2, \dots , \mu_N\}$ replacing each $\mu_i$ by 0 or 1.  We note $F_m^l$ satisfies the following equation
\begin{equation}  \label{uPPS}
  \max(F_{m+M+1}^{l+1}+F_{m-M}^l-1, F_m^l+F_{m+1}^{l+1}) = \max( F_{m+M}^{l+1}+F_{m-M+1}^l-1, F_{m+1}^l+F_{m}^{l+1} ),  
\end{equation}
which is obtained by ultradiscretizing (\ref{dPPS}).  
\section{Relation between the uPPS Equation and the uhLV Equation }
The discrete hungry Lotka-Volterra equation is ultradiscretized as
\begin{equation}  \label{uhLV}
  F_{m+1}^l+F_m^{l+1}=\max(F_m^l+F_{m+1}^{l+1}, F_{m-M}^l+F_{m+M+1}^{l+1}-1).  
\end{equation}
In addition, the uPPS solution to Eq.(\ref{uhLV}) is proposed as follows\cite{Nakamura}:  
\begin{equation}  \label{uhLV sol}
  F_m^l= \max_{\mu_i=0, 1}\Bigl( \sum_{1\le i\le N}\mu_iS_i(l, m)-(M+1)\sum_{1\le i<j\le N}\mu_i\mu_j P_j+ \sum_{1\le j\le N}\mu_j\Phi_j(m+\sum_{1\le i\le j-1}\mu_i) \Bigr),   
\end{equation}
where $S_i(l, m) = (MP_i-1)l-P_im+C_i$.  Here $P_i$ and $C_i$ are arbitrary parameters satisfying $1/M\le P_N\le \dots \le P_2\le P_1$.  Function $\Phi_i(m)$  is a periodic function satisfying $\Phi(m)=\Phi(m+M)$ and the following relations,  
\begin{equation}  \label{cond phi uhLV}
\begin{aligned}
   P_i\ge &|\Phi_i(m)-\Phi_i(m+1)|,  \\
  |P_i-P_j|\ge &|\Phi_i(m)-\Phi_i(m+1)-(\Phi_j(m)-\Phi_j(m+1))|  .   
\end{aligned}
\end{equation}
\indent We show (\ref{Nsol}) is reduced into (\ref{uhLV sol}).  Assume $P_i\ge 1/M$ for $i=1, 2, \dots , N$ in (\ref{Nsol}), then $S_i$ is rewritten as $l \max(0, MP_i-1)-mP_i+C_i= (MP_i-1)l-mP_i+C_i$.  Moreover ultradiscretization of (\ref{cond phi}) derives   
\begin{equation}  \label{cond phi2}
\begin{aligned}
  P_j-P_i > &\Phi_j(m)-\Phi_j(m-1)-(\Phi_i(m)-\Phi_i(m-1)), \\
  P_j+P_i \ge &\Phi_j(m)-\Phi_j(m-1)+\Phi_i(m+1)-\Phi_i(m), 
\end{aligned}
\end{equation}  
for $i>j$ and $m=1, 2, \dots , M$.  Thus, if we set distinct wave numbers $P_i$ and periodic functions $\Phi_i(m)$ so that they satisfy (\ref{cond phi uhLV}) and (\ref{cond phi2}), the solution (\ref{Nsol}) is reduced into (\ref{uhLV sol}).  In other words, the solution (\ref{Nsol}) under the assumptions satisfies the uhLV equation (\ref{uhLV}).  
\section{Concluding Remarks}
We derived the dPPS equation from the discrete DKP equation.  It has the dPPS solution, which presents internal oscillatory motion.  Moreover, the dPPS solution under the assumptions (\ref{cond phi}) is ultradiscretized, and we obtained the uPPS solution which is also periodically oscillating and satisfies the uPPS equation (\ref{uPPS}).  We showed that the solution  in this paper reduces to the uPPS solution to the uhLV equation in a special case.  It suggests that there may be a direct connection between DKP and KP in ultradiscrete case.  It might be interesting to clarify the relation between  integrable hierarchies and construct various soliton equations admitting periodic phase soliton solutions for both discrete and ultradiscrete systems.  We comment that the assumption (\ref{cond phi}) is not the necessary condition for taking ultradiscrete limit.  There are other choices of parameters which derive solutions for the uPPS equation.  Investigating such solutions may be a future work.  
\subsection*{Acknowledgments }
This work was supported by JSPS KAKENHI Grand Numbers JP24340029, JP26610029 and JP18H01130.
\appendix
\section{A Formula for Pfaffian}
We first show the following theorem\cite{Hirota-RIMS2012}.  
\begin{theorem} \label{theorem A}
Let $n$, $m$ be any positive integers.  Suppose 
\begin{equation}  \label{A-1}
  \pf(i, j) = \pf (d_1, d_2, \dots , d_{2m}, \alpha_i, \alpha _j)/\pf (d_1, d_2, \dots , d_{2m}), 
\end{equation}
for  $i, j=1, 2, \dots, 2n$.  Then we have
\begin{equation}  \label{A-2}
  \pf(1, 2, \dots , 2n) = \pf (d_1, d_2, \dots , d_{2m}, \alpha_1, \alpha _2, \dots , \alpha _{2n})/\pf (d_1, d_2, \dots , d_{2m}).  
\end{equation}
\end{theorem}
\textbf{Proof}. \ We prove the theorem by using mathematical induction on $n$. Assume 
\begin{equation}  \label{A-3}
  \pf(1, 2, \dots , 2n-2) = \frac{\pf (d_1, d_2, \dots , d_{2m}, \alpha_1, \alpha_2, \dots , \alpha _{2n-2})}{\pf (d_1, d_2, \dots , d_{2m})  }.  
\end{equation}
The pfaffian $\pf(1, 2, \dots, 2n)$ is obtained by the identity 
\begin{equation}
  \pf(1, 2, \dots , 2n) = \sum_{j=2}^{2n}(-1)^j\pf (1, j) \pf (2, 3, \dots , \widehat {j} , \dots 2n), 
\end{equation}
where the symbol $\widehat{j}$ means the deletion of $j$.  By using (\ref{A-1}) and (\ref{A-3}), the right-hand side is rewritten as 
\begin{equation}  \label{A-4}
\begin{aligned}
  \sum_{j=2}^{2n}(-1)^j\frac{\pf (d_1, d_2, \dots , d_{2m}, \alpha _1, \alpha_j)}{\pf(d_1, d_2, \dots , d_{2m})}  \frac{\pf (d_1, d_2, \dots , d_{2m}, \alpha_2, \alpha_3, \dots , \widehat {\alpha _j} , \dots , \alpha _{2n})}{\pf(d_1, d_2, \dots , d_{2m})},  
\end{aligned}
\end{equation}
and moreover we have
\begin{equation}  
\begin{aligned}
  &\sum_{j=2}^{2n}(-1)^j\pf (d_1, d_2, \dots , d_{2m}, \alpha _1, \alpha_j)\pf (d_1, d_2, \dots , d_{2m}, \alpha_2, \alpha_3, \dots , \widehat {\alpha _j} , \dots , \alpha _{2n})\\
  =& \pf(d_1, d_2, \dots , d_{2m})\pf(d_1, d_2, \dots , d_{2m}, \alpha_1, \alpha_2, \dots , \alpha_{2n}),  
\end{aligned}
\end{equation}
due to the Pl\"ucker relation of pfaffian type\cite{DIS}.  Therefore we obtain (\ref{A-2}).  
\section{Proof of the Discrete DKP Equation (\ref{DKP})}
We show the pfaffian (\ref{DKP sol1}) satisfies (\ref{DKP}) by using Theorem \ref{theorem A}.  In this appendix, we denote $(k, l, m)$ as $(k_1, k_2, k_3)$, and  $\tau(k) =\tau (k_1, k_2, k_3)=\pf(\alpha_1, \dots , \alpha_N, \beta_N, \dots , \beta_1)_{k_1, k_2, k_3}$.  Moreover we use the following notations, 
\begin{equation}
\begin{aligned}
  \tau(k_1) =&\tau (k_1+1, k_2, k_3)=\pf(\alpha_1, \alpha_2, \dots , \alpha_N, \beta_N, \dots , \beta_2, \beta_1)_{k_1+1, k_2, k_3}, \\
  \tau(\widetilde {k_1}) =&\tau (k_1, k_2+1, k_3+1)=\pf(\alpha_1, \alpha_2, \dots , \alpha_N, \beta_N, \dots , \beta_2, \beta_1)_{k_1, k_2+1, k_3+1}, \\
  \tau(\overline {k}) =&\tau (k_1+1, k_2+1, k_3+1)=\pf(\alpha_1, \alpha_2, \dots , \alpha_N, \beta_N, \dots , \beta_2, \beta_1)_{k_1+1, k_2+1, k_3+1},  
\end{aligned}
\end{equation}
and $\tau (k_2)$, $\tau (k_3)$, $\tau (\widetilde{k_2})$, $\tau (\widetilde{k_3})$ are similarly defined.  Let us introduce new indices, $d_1$, $d_2$, $d_3$ and $d_4$, which are defined by 
\begin{equation}
\begin{aligned}
  &\pf(d_0, d_1) =\pf(d_0, d_2)=1, \quad \pf(d_1, d_2) =-\delta, \\
  &\pf(d_0, d_3)=\pf(d_1, d_3)=\pf(d_2, d_3)=-1.  
\end{aligned}
\end{equation}
In these notations, the discrete DKP equation (\ref{DKP}) is expressed by 
\begin{equation}   \label{B-1}
\begin{aligned}
  &\pf(d_0, d_1, d_2, d_3)\tau (\overline{k})\tau (k)  =\pf(d_0, d_1)\pf(d_2, d_3)\tau(k_1)\tau(\widetilde{k_1})\\
  &-\pf(d_0, d_2)\pf(d_1, d_3)\tau(k_2)\tau(\widetilde{k_2})+\pf(d_0, d_3)\pf(d_1, d_2)\tau(k_3)\tau(\widetilde{k_3}).
\end{aligned}
\end{equation}
 \par
Now let us denote the $(i, j)$-elements of pfaffians $\tau(k_l)$, $\tau(\widetilde{k_l})$, $\tau(\overline{k})$ as $\tau_{ij}(k_l)$, $\tau_{ij}(\widetilde{k_l})$, $\tau_{ij}(\overline{k})$, respectively, where $l=1, 2, 3$ and $i, j=\alpha_1, \alpha_2, \dots , \alpha_N, \beta_1, \beta_2, \dots , \beta_N$.  Then by defining pfaffian elements, 
\begin{equation}
\begin{aligned}
  &\pf(d_0, \alpha_h) =\pf(d_3, \alpha_h)=\eta_h(k_1, k_2), \quad \pf(d_1, \alpha_h) =q_h\eta_h(k_1, k_2), \quad \pf(d_2, \alpha_h) =\omega _h\eta_h(k_1, k_2), \\
  & \pf(d_0, \beta_h) = \pf(d_1, \beta_h) =\pf(d_2, \beta_h) =\psi _h(k_3), \quad \pf(d_3, \beta_h) =\psi _h(k_3+1),  \qquad (1\le h \le N),   
\end{aligned}
\end{equation}
we obtain 
\begin{equation}  \label{B-2}
\begin{aligned}
  \tau_{ij}(k_l) =&\pf(d_0, d_l, i, j)/\pf(d_0, d_l), \\
  \tau_{ij}(\widetilde{k_l}) =&\pf(d_1, \widehat{d_l}, d_3, i, j)/\pf(d_1, \widehat{d_l}, d_3), \\
  \tau_{ij}(\overline{k}) =&\pf(d_0, d_1, d_2, d_3, i, j)/\pf(d_0, d_1, d_2, d_3), 
\end{aligned}
\end{equation}
for $l=1, 2, 3$ and $i, j=\alpha_1, \alpha_2, \dots , \alpha_N, \beta_1, \beta_2, \dots , \beta_N$.  For example, in the case of $l=1$, $i=\alpha_i$, $j=\alpha_j$, we have
\begin{equation}\label{B-2a}
\begin{aligned}
  \tau _{\alpha_i \alpha_j}(k_1) =&\pf(\alpha_i, \alpha_j) _{k_1+1, k_2, k_3} =a_{ij} q_iq_j \eta_i(k_1, k_2)\eta_j(k_1, k_2)=\pf(d_0, d_1, \alpha_i, \alpha_j)/\pf(d_0, d_1), \\
  \tau _{\alpha_i \alpha_j}(\widetilde{k_1}) =&\pf(\alpha_i, \alpha_j) _{k_1, k_2+1, k_3+1} =a_{ij} \omega_i\omega_j \eta_i(k_1, k_2)\eta_j(k_1, k_2)=\pf(d_2, d_3, \alpha_i, \alpha_j)/\pf(d_2, d_3), \\
  \tau _{\alpha_i \alpha_j}(\overline{k}) =&\pf(\alpha_i, \alpha_j) _{k_1+1, k_2+1, k_3+1}   =a_{ij} q_iq_j\omega _i\omega_j \eta_i(k_1, k_2)\eta_j(k_1, k_2)\\
  =&\pf(d_0, d_1, d_2, d_3, \alpha_i, \alpha_j)/\pf(d_0, d_1, d_2, d_3).  
\end{aligned}
\end{equation}
The other cases of (\ref{B-2}) can be proved in a similar way.  Then due to Theorem \ref{theorem A}, the pfaffians are given by 
\begin{equation}  \label{B-3}
\begin{aligned}
  \tau(k_l) =&\pf(d_0, d_l, \alpha_1, \alpha_2, \dots , \alpha_N, \beta_N, \dots , \beta_2, \beta_1)/\pf(d_0, d_l),\\
  \tau(\widetilde{k_l}) =&\pf(d_1, \widehat{d_l}, d_3, \alpha_1, \alpha_2, \dots , \alpha_N, \beta_N, \dots , \beta_2, \beta_1)/\pf(d_1, \widehat{d_l}, d_3), \\
  \tau(\overline{k}) =&\pf(d_0, d_1, d_2, d_3, \alpha_1, \alpha_2, \dots , \alpha_N, \beta_N, \dots , \beta_2, \beta_1)/\pf(d_0, d_1, d_2, d_3),  
\end{aligned}
\end{equation}
and (\ref{B-1}) is rewritten as
\begin{equation}
\begin{aligned}
  &\pf(d_0, d_1, d_2, d_3, \alpha_1, \alpha_2, \dots , \alpha_N, \beta_N, \dots , \beta_2, \beta_1)\pf(\alpha_1, \alpha_2, \dots , \alpha_N, \beta_N, \dots , \beta_2, \beta_1)\\
  =&\pf(d_0, d_1, \alpha_1, \alpha_2, \dots , \alpha_N, \beta_N, \dots , \beta_2, \beta_1)\pf(d_2, d_3, \alpha_1, \alpha_2, \dots , \alpha_N, \beta_N, \dots , \beta_2, \beta_1)\\
  -&\pf(d_0, d_2, \alpha_1, \alpha_2, \dots , \alpha_N, \beta_N, \dots , \beta_2, \beta_1)\pf(d_1, d_3, \alpha_1, \alpha_2, \dots , \alpha_N, \beta_N, \dots , \beta_2, \beta_1)\\
  +&\pf(d_0, d_3, \alpha_1, \alpha_2, \dots , \alpha_N, \beta_N, \dots , \beta_2, \beta_1)\pf(d_1, d_2, \alpha_1, \alpha_2, \dots , \alpha_N, \beta_N, \dots , \beta_2, \beta_1),  
\end{aligned}
\end{equation}
which is nothing but the pfaffian's identity\cite{Hirota-direct}.  This completes the proof.  
\section{Proof of (\ref{rel pf})}
In this appendix, we prove the relation (\ref{rel pf}) by showing the correspondence of the terms which have $s_1s_2\dots s_n$ in the both sides.  When $n$ is even,  the terms on the right hand side of (\ref{rel pf}) are given by $\pf(a_1, a_2, \dots, a_n)\pf(b_n, \dots , b_2, b_1)$ since 
\begin{equation}  
\begin{aligned}
\pf(a_1, a_2, \dots, a_n)=&
  \sum \nolimits^{\prime} \text{sgn} \begin{pmatrix} 
  1 & 2 & \dots & n \\
  i_1 & i_2 & \dots & i_n
\end{pmatrix} \pf(a_{i_1}, a_{i_2})\pf(a_{i_3}, a_{i_4})\dots \pf(a_{i_{n-1}}, a_{i_n})\\
=&  \sum\nolimits^{\prime} \text{sgn} \begin{pmatrix} 
  1 & 2 & \dots & n \\
  i_1 & i_2 & \dots & i_n
\end{pmatrix} a_{i_1 i_2}a_{i_3 i_4}\dots a_{i_{n-1} i_n} s_1s_2\dots s_n, 
\end{aligned}
\end{equation}
where $\sum '$ denotes the summation over all permutations $\{i_1, i_2, \dots , i_n\}$ of $\{1, 2, \dots , n\}$ which satisfy the inequalities
\begin{equation}
\begin{aligned}
  &i_1<i_2, \ i_3<i_4, \ \dots , \ i_{n-1}<i_n,  \qquad \text{and} \qquad i_1<i_3<\dots <i_{n-1}.
\end{aligned}
\end{equation}
Let us consider the terms on the left hand side of (\ref{rel pf}).   We can find that all of the terms which have $s_1s_2\dots s_n$ belong to 
\begin{equation}
  \pf(a_1, a_2, \dots , a_n, b_n, \dots , b_2, b_1)\pf(a_{N-n}, b_{N-n})\pf(a_{N-n+1}, b_{N-n+1})\dots \pf(a_N, b_N)
\end{equation}
since $\pf(a_i, b_j)$ or $\pf(a_i ,a _j)$ has $s_i\phi _j$ or $a_{ij}s_is_j$ for $i\not=j$.  Then we consider 
\begin{equation}  \label{C-1}
  \pf(a_1, a_2, \dots , a_n, b_n, \dots , b_2, b_1).     
\end{equation}
Expansion of (\ref{C-1}) has the terms  
\begin{equation}  \label{C-2}
\begin{aligned}
  & (-1)^{i_1+i_2+j_1+j_2}\pf(a_{i_1}, b_{j_2})\pf(a_{i_2}, b_{j_1})\pf(\bullet), \\   
  &(-1)^{i_1+i_2+j_1+j_2+1}\pf(a_{i_1}, b_{j_1})\pf(a_{i_2}, b_{j_2})\pf(\bullet), 
\end{aligned}
\end{equation}
where $1\le i_1<i_2\le n$, $1\le j_1<j_2\le n$, and $\pf(\bullet )$ denotes 
\begin{equation}
  \pf(a_1, \dots , \widehat{a_{i_1}}, \dots , \widehat{a_{i_2}},\dots , a_n, b_n, \dots , \widehat{b_{j_2}}, \dots , \widehat{b_{j_1}}, \dots , b_1). 
\end{equation}
Each of (\ref{C-2}) has $s_{i_1}s_{i_2}$, however, the sum of them does not since   
\begin{equation}
\begin{aligned}
  &\pf(a_{i_1}, b_{j_2})\pf(a_{i_2}, b_{j_1})\pf(\bullet )-\pf(a_{i_1}, b_{j_1})\pf(a_{i_2}, b_{j_2})\pf(\bullet )  \\
  =&(\delta_{i_1, j_2}\delta_{i_2, j_1}+\delta_{i_1, j_2}s_{i_2}\phi_{j_1}+\delta_{i_2, j_1}s_{i_1}\phi_{j_2}-\delta_{i_1, j_1}\delta_{i_2, j_2}-\delta_{i_1, j_1}s_{i_2}\phi_{j_2}-\delta_{i_2, j_2}s_{i_1}\phi_{j_1})\pf(\bullet ).  
\end{aligned}
\end{equation}
Thus, all the terms of $s_1 s_2\dots s_n$ among the expansion of (\ref{C-1}) are expressed by 
\begin{equation}  \label{C-3}
\begin{aligned}
  \sum \nolimits^{\prime} \text{sgn} \begin{pmatrix} 
  1 & 2 & \dots & n \\
  i_1 & i_2 & \dots & i_n
\end{pmatrix} \pf(a_{i_1}, a_{i_2})\pf(a_{i_3}, a_{i_4})\dots \pf(a_{i_{n-1}}, a_{i_n})\\
\times   \sum \nolimits^{\prime} \text{sgn} \begin{pmatrix} 
  1 & 2 & \dots & n \\
  j_n & j_{n-1} & \dots & j_1
\end{pmatrix} \pf(b_{j_n}, b_{j_{n-1}})\dots \pf(b_{j_4}, b_{j_3})\pf(b_{j_2}, b_{j_1})  ,
\end{aligned}
\end{equation}
and it corresponds to $\pf(a_1, a_2, \dots, a_n)\pf(b_n, \dots , b_2, b_1)$.  \par
When $n$ is odd, the terms of $s_1s_2\dots s_n$ on the right hand side of (\ref{rel pf}) are given by $\pf(a_1, a_2, \dots, a_n, \emptyset)\pf(\emptyset, b_n, \dots , b_2, b_1) $.  On the other hand, the terms on the left hand side are obtained from 
\begin{equation}
  \sum _{1\le i\le n}\sum _{1\le j\le n}(-1)^{i+j}\pf(a_i, b_j) \pf(a_1, \dots , \widehat{a_i}, \dots, a_n, b_n, \dots , \widehat{b_j}, \dots , b_1),  
\end{equation}
among which the terms containing $s_1s_2\dots s_n$ appear in
\begin{equation}
\begin{aligned}
  &\sum _{1\le i\le n}\sum _{1\le j\le n}(-1)^{i+j} s_i\phi_j(m) \pf(a_1, \dots , \widehat{a_i}, \dots, a_n)\pf(b_n, \dots , \widehat{b_j}, \dots , b_1)\\
  =&\sum _{1\le i\le n}\sum _{1\le j\le n} (-1)^{i+j}\pf(a_i, \emptyset )\pf(\emptyset, b_j) \pf(a_1, \dots , \widehat{a_i}, \dots, a_n)\pf(b_n, \dots , \widehat{b_j}, \dots , b_1)\\
  =&\pf(a_1, a_2, \dots, a_n, \emptyset)\pf(\emptyset, b_n, \dots , b_2, b_1).  
\end{aligned}
\end{equation}
Therefore, the proof is done.  
\section{Expansion of $\pf(b_N, \dots , b_2, b_1)$}
We consider the expansion of $\pf(b_N, \dots , b_2, b_1)$ and its greatest term.  We use a notation 
\begin{equation}  \label{def Gamma}
  \Gamma _{ij}^\mu = p_i^{\mu }p_j^{\mu -1}\phi_i(m+\mu)\phi_j(m+\mu-1).  
\end{equation} 
Then, we get $\Gamma_{ij}^{\mu }>\Gamma_{ji}^{\mu }$ for $i>j$ and $\Gamma_{ij}^{\mu }\ge \Gamma_{ij}^{\nu }$ for $\mu <\nu$ from (\ref{cond phi}).  Moreover the followings also hold.  
\begin{equation}  \label{D-1}
\begin{aligned}
  &\Gamma _{i_1i_2} ^{\mu}\Gamma _{i_3i_4} ^{\mu}-\Gamma _{i_1i_4} ^{\mu}\Gamma _{i_3i_2} ^{\mu}=0, \\
  &\Gamma _{i_1i_2} ^{\mu+1}\Gamma _{i_3i_4} ^{\mu}-\Gamma _{i_1i_3} ^{\mu+1}\Gamma _{i_2i_4} ^{\mu}=0, \\
  &\phi_{i_1}(m)\Gamma _{i_2i_3} ^1-\phi_{i_3}(m)\Gamma _{i_2i_1} ^1=0,    
\end{aligned}
\end{equation}
for $i_1, i_2, i_3, i_4= 1,2, \dots, N$ and $\mu=1, 2, \dots $.  Using this notation, $\pf (b_i, b_j)$ can be expressed by 
\begin{equation}
\begin{aligned}
  \pf (b_i, b_j) =&  \frac{1}{1-p_i^Mp_j^M} \sum_{\mu=1}^M (p_i^{\mu }p_j^{\mu -1}\phi_i(m+\mu)\phi_j(m+\mu-1)-p_j^{\mu }p_i^{\mu -1}\phi_j(m+\mu)\phi_i(m+\mu-1))  \\
  =&  \sum _{\nu =0}^{\infty } p_i^{\nu M}p_j^{\nu M}\sum_{\mu=1}^M (p_i^{\mu }p_j^{\mu -1}\phi_i(m+\mu)\phi_j(m+\mu-1)-p_j^{\mu }p_i^{\mu -1}\phi_j(m+\mu)\phi_i(m+\mu-1)) \\
  =& \sum_{\mu=1}^{\infty} (p_i^{\mu }p_j^{\mu -1}\phi_i(m+\mu)\phi_j(m+\mu-1)-p_j^{\mu }p_i^{\mu -1}\phi_j(m+\mu)\phi_i(m+\mu-1)) \\
  =& \sum_{\mu=1}^{\infty} (\Gamma _{ij}^{\mu }-\Gamma _{ji}^{\mu})  
\end{aligned}
\end{equation}
since $p_i<1$ and $\phi_i(m+M)=\phi_i(m)$.  When $N$ is even, we obtain
\begin{equation}  \label{D-2}
\begin{aligned}
  &\pf(b_N, b_{N-1}, \dots, b_2, b_1) \\
  =&(-1)^{N/2}\pf(b_1, b_2, \dots, b_N)\\
  =& (-1)^{N/2}\sum \nolimits^{\prime}\text{sgn} \begin{pmatrix} 
  1 & 2 & \dots & N \\
  i_1 & i_2 & \dots & i_N
\end{pmatrix}
\sum _{\mu_1=1}^{\infty}(\Gamma _{i_1 i_2}^{\mu_1} -\Gamma _{i_{2} i_1}^{\mu_{1}})\sum _{\mu_{3}=1}^{\infty}(\Gamma _{i_{3} i_{4}}^{\mu_{3}} -\Gamma _{i_4i_3}^{\mu_{3}})\\
  &  \dots \sum _{\mu_{N-1}=1}^{\infty}(\Gamma _{i_{N-1}i_N}^{\mu_{N-1}} -\Gamma _{i_{N} i_{N-1}}^{\mu_{N-1}}) \\
  =& (-1)^{N/2}\sum _{\mu_{1}, \mu_{3}, \dots , \mu_{N-1} =1}^{\infty}\sum \nolimits^{\prime \prime}\text{sgn} \begin{pmatrix} 
  1 & 2 & \dots & N \\
  i_1 & i_2 & \dots & i_N
\end{pmatrix}\Gamma_{i_{1}i_{2}}^{\mu_1}\Gamma_{i_{3}i_{4}}^{\mu_{3}}\dots \Gamma_{i_{N-1}i_N}^{\mu_{N-1}},  
\end{aligned}
\end{equation}
where $\sum ''$ denotes the summation over all permutations $\{i_1, i_2, \dots, i_N\}$ of $\{1, 2, \dots, N\}$ which satisfy the inequalities $i_1<i_3<\dots <i_{N-1}$.  From (\ref{D-1}), the terms which have $\Gamma _{i_1i_2}^{\mu }\Gamma _{i_3i_4}^{\mu }$ or $\Gamma _{i_1i_2}^{\mu +1}\Gamma _{i_3i_4}^{\mu }$ in (\ref{D-2}) vanish for $\mu= 1, 2, \dots $ and $i_1, i_2, i_3, i_4=1, 2, \dots , N$.  In addition, $\Gamma ^{\mu}_{i_2i_1}\Gamma ^{\mu+2}_{i_4i_3}> \Gamma ^{\mu}_{i_4i_3}\Gamma ^{\mu+2}_{i_2i_1}$ holds for $i_1<i_2<i_3<i_4$ and $\mu=1, 2, 3, \dots $ from (\ref{cond phi}). Due to these facts,  the greatest term in (\ref{D-2}) is given as $\Gamma _{21}^1\Gamma _{43}^3\dots \Gamma _{N, N-1}^{N-1}=\prod _{i=1}^Np_i^{i-1}\phi_i(m+i-1)$.  When $N$ is odd, 
\begin{equation}  \label{D-4}
\begin{aligned}
  &\pf(b_N, \dots , b_2, b_1)\\
  =& (-1)^{(N-1)/2}\pf(\emptyset, b_1, b_2, \dots , b_N)\\
  = (-1)^{(N-1)/2}&\sum _{\mu_2, \mu_4, \dots , \mu_{N-1} =1}^{\infty}\sum \nolimits^{\prime \prime }\text{sgn} \begin{pmatrix} 
  1 & 2 & \dots & N \\
  i_1 & i_2 & \dots & i_N
\end{pmatrix}\phi_{i_1}\Gamma_{i_2i_3}^{\mu_2}\dots \Gamma_{i_{N-1}i_N}^{\mu_{N-1}} .  
\end{aligned}
\end{equation}
From (\ref{D-1}), the terms which have $\phi_{i_1}\Gamma _{i_2i_3}^{1}$ also vanish for $i_1, i_2, i_3=1, 2, \dots , N$.  Thus the greatest term is given as $\phi_1(m)\Gamma _{32}^2\Gamma _{54}^4\dots \Gamma _{N, N-1}^{N-1}=\prod _{i=1}^Np_i^{i-1}\phi_i(m+i-1)$.  

\end{document}